\documentclass{article}
\usepackage[final,sglblindworkshop]{neurips_2026}
\workshoptitle{The 1st Workshop on Interpreting Agent Behavior (IAB)}
\usepackage{amsmath,amssymb,booktabs,graphicx,xcolor,microtype,url,flafter,float,tikz}
\usetikzlibrary{arrows.meta,positioning,fit,backgrounds}
\usepackage[utf8]{inputenc}
\usepackage[T1]{fontenc}
\newcommand{\Actonomy}{Act\ensuremath{\cdot}onomy}
\newcommand{\Ngames}{298{,}357}

\newcommand{\LiveGainB}{+2.52}
\newcommand{\LiveLoB}{-13.11}
\newcommand{\LiveHiB}{+18.61}

\newcommand{\LiveGainN}{+1.90}
\newcommand{\LiveLoN}{-8.38}
\newcommand{\LiveHiN}{+12.23}

\newcommand{\LiveGainP}{-1.51}
\newcommand{\LiveLoP}{-12.98}
\newcommand{\LiveHiP}{+10.20}

\newcommand{\VTwoGain}{+7.31}
\newcommand{\VTwoLo}{-0.43}
\newcommand{\VTwoHi}{+15.02}

\newcommand{\VTwoEqualGain}{+5.91}
\newcommand{\VTwoEqualLo}{-6.79}
\newcommand{\VTwoEqualHi}{+18.43}

\newcommand{\VTwoDivergence}{57.6}
\newcommand{\VTwoTextDivergence}{34.4}
\newcommand{\VTwoActionDivergence}{23.1}

\usepackage{hyperref}

\title{How Behavioral Structure Shapes Agent Strategy\\
in Language-Based Economic Games}

\author{
\textbf{Yichi Zhang}$^{1}$ \quad
\textbf{Shuo Yang}$^{1}$ \quad
\textbf{Jun Wang}$^{2}$\thanks{This work is not related to Amazon.} \quad
\textbf{Chunhua Liu}$^{1}$\\
\normalfont $^{1}$University of Melbourne \qquad $^{2}$Amazon\\
\small\texttt{\{yzhang3504,syang4418\}@student.unimelb.edu.au} \\\quad
\small\texttt{junwc@amazon.com} \\ \quad
\texttt{chunhua.liu1@unimelb.edu.au}
}
\begin{document}
\maketitle

\begin{abstract}
In competitive multi-agent economic games, agents continuously interact, strategize, and make decisions within a shared market. Consequently, every message produced by a language agent changes both the current economic state of the market and the opponent's subsequent behavior. Although terminal rewards can indicate whether a decision was effective and broad behavioral labels can provide a coarse interpretation of the interaction, neither specifies the concrete response that should be taken next. We therefore ask whether explicitly providing an opponent's operational structure to a focal agent changes the value of its task-specific executable response strategies and the resulting game outcome. Using Games in Language-based Economic Environments (GLEE) as a controlled environment, we investigate this question across bargaining, negotiation, and persuasion games. We analyze \Ngames{} trajectories using seven relational profiles and decompose each task's dominant profile into ten \Actonomy{} structures, finding that dominant profiles have task-specific operational compositions. We then compare structure-blind and structure-aware policies in matched offline simulation and prospective online play. In fixed $334$-game online cohorts for each task, the aware-minus-blind official-percentile point estimates of the profile-gated eligible samples are \LiveGainB{}, \LiveGainN{}, and \VTwoEqualGain{} points in bargaining, negotiation, and persuasion, respectively (corresponding $95\%$ confidence intervals include zero). Taken together, the results distinguish strategy selection from response realization: structure re-ranks strategies, while in persuasion the online point estimate becomes positive only when the policy also changes the submitted message or purchase action.
\end{abstract}

\section{Introduction}
Language-based economic games are multi-turn strategic interactions in which agents use natural language together with payoff-bearing actions, including offers, trades, and purchase decisions, to allocate resources or influence outcomes. Earlier negotiation environments and language-mediated games such as Diplomacy have explored how agents coordinate and compete through dialogue in separate settings~\citep{fair2022diplomacy,davidson2024agency}. Games in Language-based Economic Environments (GLEE) extends this paradigm by bringing bargaining, negotiation, and persuasion into a common controlled environment~\citep{shapira2026glee}. To compete in these games, large language model (LLM) agents must infer adaptive opponents, manage mixed incentives, and act over multiple turns~\citep{zhou2024sotopia,davidson2024agency}, requiring both social reasoning and strategic planning. Therefore, understanding LLM agents' behavior in these games is important for developing more reliable and effective LLM agents capable of operating in open-ended, decentralized markets.

The key challenge is that these games are inherently dynamic: interactions between focal agents and their opponents evolve over time, continually shaping game outcomes. A competent focal agent must therefore understand the current game state and anticipate how opponents may adapt their strategies in response to the focal agent's actions~\citep{fair2022diplomacy,shapira2026glee}. While prior work has explored the performance of LLMs in these games, it has largely focused on \textbf{static opponent assumptions} or \textbf{scalar terminal rewards}~\citep{liu2024agentbench}. It remains underexplored how agents should respond to different opponent profiles and whether the operational structure expressed within a profile changes strategy selection and outcomes in dynamic environments.

In this work, we investigate the role of opponent profiles and their structures in multi-turn economic games. Specifically, we characterize how LLM agents respond to different profiles and test whether modeling the structure expressed within these profiles changes strategy selection and game outcomes. We distinguish an opponent \emph{profile}---the observable relational pattern expressed in the interaction---from its \emph{operational structure}---the \Actonomy{} operations evidenced in how that pattern is produced through the opponent's decisions and actions~\citep{gao2026actonomy}. These operations are identified from observable trajectory evidence rather than interpreted as latent psychological states.
To this end, we (1) introduce a trajectory-based framework that maps a game trajectory to an opponent profile; (2) analyze an agent’s \emph{structural behavior} through the operations underlying its decisions and actions such as grounding, memory, planning, reasoning, etc; and (3) evaluate whether access to this structure changes the focal agent's selected strategy, delivered response, and resulting game outcome (e.g., allocated share, trade utility, or purchase outcome).

To obtain game trajectories with diverse opponent profiles, we deployed five LLM-based focal agents on the GLEE platform~\citep{shapira2026glee,gleecompetition2026}, where they competed against heterogeneous opponents across three economic games: bargaining, negotiation, and persuasion. Over 15 calendar days, we collected $\Ngames$ interaction trajectories. We use these trajectories to study the relationship between opponent profiles and focal-agent behavior. We then conduct two experiments to examine whether access to operational structure affects strategy selection and game outcomes: an \emph{offline matched simulation} comparing structure-blind and structure-aware policies in identical simulated contexts under common random outcomes, and a prospective \emph{online study} comparing policies with and without access to pre-response structure during live interactions in GLEE.
The primary findings of our work are as follows:
\begin{enumerate}
    \item \textbf{The same opponent profile can have different structures across games.}
    Competitive--hardline is the dominant opponent profile in both bargaining
    and negotiation, but it is expressed through different operations in the
    two games. Persuasion instead has a dominant assertive profile with a more
    message-mediated structure.

    \item \textbf{Structural information changes strategy choice, but not
    always for the better.}
    In the offline simulation, the structure-aware policy selects different
    strategies from the structure-blind policy in all three games. However,
    its average score improves only in negotiation, decreases slightly in
    bargaining, and remains approximately unchanged in persuasion.

    \item \textbf{Performance changes depend on whether structure changes the
    agent's actual response.}
    In the online study, bargaining and negotiation have generally positive point
    estimates when structure changes strategy routing. In persuasion, routing
    alone produces a negative estimate. The estimate becomes positive when the
    structure-aware policy can also change the message or purchase action
    submitted to the game.
\end{enumerate}

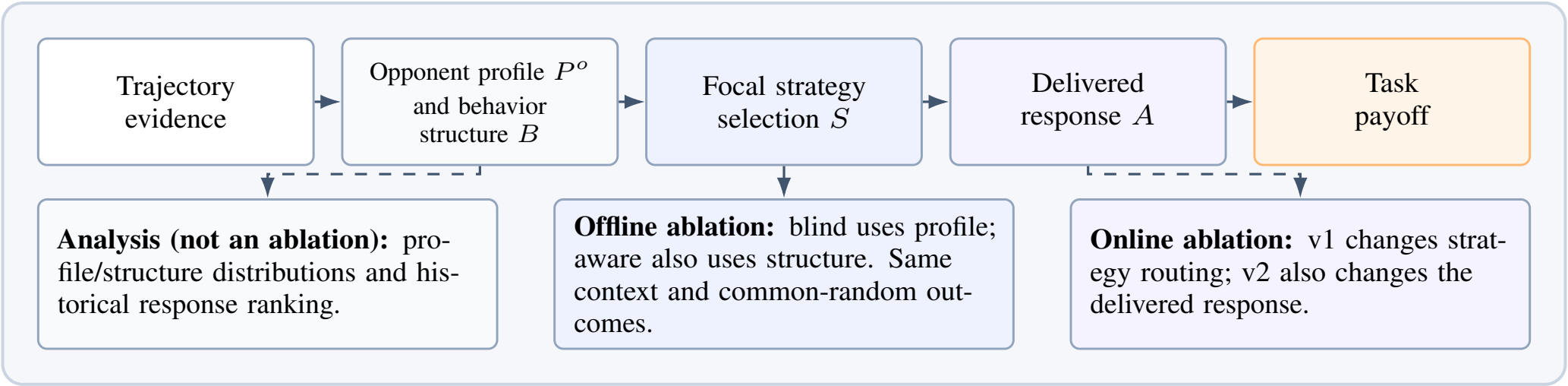
\begin{figure}[t]
\centering
\definecolor{wfouter}{HTML}{F5F7FB}
\definecolor{wfouterborder}{HTML}{CBD5E1}
\definecolor{wfborder}{HTML}{91A4BC}
\definecolor{wfedge}{HTML}{42556E}
\definecolor{wftop}{HTML}{FFFFFF}
\definecolor{wfanalysis}{HTML}{F8FAFC}
\definecolor{wfoffline}{HTML}{EEF2FF}
\definecolor{wfonline}{HTML}{F5F3FF}
\definecolor{wfpayoff}{HTML}{FFF4E8}
\begin{tikzpicture}[
  >=Latex,
  box/.style={draw=wfborder, line width=.55pt, rounded corners=2.2pt,
              align=center, minimum width=2.4cm, minimum height=11mm,
              text width=2.06cm,
              font=\scriptsize, fill=wftop},
  study/.style={draw=wfborder, line width=.55pt, rounded corners=2.2pt,
                align=left, minimum width=4cm, minimum height=13mm,
                text width=3.66cm,
                font=\scriptsize},
  arrow/.style={-{Latex[length=1.8mm,width=1.25mm]},
                line width=.6pt, draw=wfedge}
]
\node[box] (traj) at (1.2,0) {Trajectory\\evidence};
\node[box, fill=wfanalysis] (repr) at (3.85,0)
{\scalebox{.9}{\shortstack{Opponent profile $P^o$\\and behavior\\structure $B$}}};
\node[box, fill=wfoffline] (strategy) at (6.5,0) {Focal strategy\\selection $S$};
\node[box, fill=wfonline] (response) at (9.15,0) {Delivered\\response $A$};
\node[box, fill=wfpayoff, draw=orange!55] (payoff) at (11.8,0) {Task\\payoff};
\draw[arrow] (traj) -- (repr);
\draw[arrow] (repr) -- (strategy);
\draw[arrow] (strategy) -- (response);
\draw[arrow] (response) -- (payoff);

\node[study, fill=wfanalysis] (analysis) at (2,-1.5)
{\textbf{Analysis (not an ablation):} profile/structure distributions and historical response ranking.};
\node[study, fill=wfoffline] (offline) at (6.5,-1.5)
{\textbf{Offline ablation:} blind uses profile; aware also uses structure. Same context and common-random outcomes.};
\node[study, fill=wfonline] (online) at (11,-1.5)
{\textbf{Online ablation:} v1 changes strategy routing; v2 also changes the delivered response.};
\draw[arrow, dashed] (repr.south) -- ++(0,-.7mm) -| (analysis.north);
\draw[arrow, dashed] (strategy.south) -- ++(0,-2mm) -| (offline.north);
\draw[arrow, dashed] (response.south) -- ++(0,-.7mm) -| (online.north);
\begin{scope}[on background layer]
\node[
  fit=(traj)(repr)(strategy)(response)(payoff)(analysis)(offline)(online),
  draw=wfouterborder,
  fill=wfouter,
  rounded corners=5pt,
  line width=.75pt,
  inner xsep=3mm,
  inner ysep=3mm
] {};
\end{scope}
\end{tikzpicture}
\caption{\textbf{Study framework.} The solid path gives the proposed
profile--structure--strategy mechanism and ends at task payoff. The lower boxes
distinguish descriptive analysis from the offline and online experiments and show
the information available in the structure-blind and structure-aware conditions.}
\label{fig:overall-framework}
\end{figure}

\section{Related Work}
\paragraph{Language agents in economic games.}
Language-mediated economic games evaluate strategic decision making under
multi-turn interaction. Negotiation environments study how LLM agents bargain
across models and settings~\citep{davidson2024agency}, while Diplomacy combines
natural-language communication with consequential coordination and competition~\citep{fair2022diplomacy}. Interactive benchmarks further evaluate social
reasoning and sequential action~\citep{zhou2024sotopia,liu2024agentbench}.
GLEE places bargaining, negotiation, and persuasion in a common controlled environment~\citep{shapira2026glee}; we use it to examine whether a finer representation of opponent behavior changes the focal agent's strategy and delivered response. However, prior evaluations commonly rely on fixed interaction settings or summarize performance with scalar terminal outcomes. They therefore provide limited evidence about whether representing an adaptive opponent changes the focal agent's selected strategy and delivered response~\citep{liu2024agentbench,ma2024agentboard}.

\paragraph{Opponent and user modeling.}
Prior work represents counterpart behavior through strategic types,
interaction patterns, or interpretable agent traces~\citep{pruitt1981negotiation,ma2024agentboard,cemri2025fail}. The dual-concern
tradition characterizes cooperative, accommodating, assertive, and competitive
relations~\citep{pruitt1981negotiation}, while work on social power and
deceptive communication motivates separate coercive and manipulative patterns
\citep{french1959bases,raven1965social,scheurer2024deception}. Building on these
lines, we use an evidence-bounded opponent-profile taxonomy and the
\Actonomy{} vocabulary for observable operational structure
\citep{gao2026actonomy}. These representations describe trajectory evidence;
they are not interpreted as latent psychological traits.

\section{Method}
Aiming to understand how LLM agents behave across different opponent profiles and how access to structure-related information affects focal-agent decisions, we first describe the focal-agent design and trajectory collection
(\S\ref{ssec:agent_setup}), followed by opponent-profile extraction and behavior
analysis (\S\ref{ssec:trajectory_analysis}). We then evaluate how structure access
affects focal-agent behavior, strategy selection, generated responses, and final game outcomes (payoffs)
(\S\ref{ssec:structure_evaluation}). Figure~\ref{fig:overall-framework} summarizes
the study framework.

\subsection{Agent Setup and Data Collection}
\label{ssec:agent_setup}
\paragraph{Focal-Agent Design.}
We developed five focal agents to make legal, task-specific decisions under GLEE's interaction and action constraints. Each agent shares the same four-step pipeline, consisting of defensive state parsing, task-specific strategy selection, exact legal-action validation, and deterministic fallback. The deployed suite
combines deterministic economic policies with a rate-limited Claude (Sonnet 5) decision route; when a model output is unavailable, invalid, or exceeds the decision budget, a deterministic policy supplies a legal action. During development, we used human-in-loop development to iterate the task-specific policies and fixed the evaluated variants before the reported analyses. The language-model route conditions on the parsed task state and dialogue context and proposes a task-valid candidate, which passes through the same validator as deterministic candidates. Development was author-guided rather than self-evolving: policies were revised before evaluation, then frozen, and no reported evaluation outcome updated them online. Appendix~\ref{app:agent-design} provides further implementation details and an overview of the development and runtime phases (Figure~\ref{fig:agent-design}). 


\paragraph{Trajectory Collection.}
We conduct our empirical investigation on the GLEE platform~\citep{shapira2026glee,gleecompetition2026}, which provides standardized language-based economic games with server-generated configurations and complete interaction logs. Over 15 days, five focal agents collected $\Ngames$ complete game trajectories comprising $3{,}144{,}076$ decisions across bargaining, negotiation, and persuasion. GLEE converts each task-specific payoff into an opponent-adjusted percentile $q_g\in[0,1]$ within the same configuration and role. We analyze the normalized online score $Y_g=100q_g$, so reported online contrasts are official percentile-point differences.\footnote{Task-level corpus counts, agent configurations, fallback behavior, and the score transformation and historical rating-update definition are presented in Appendix~\ref{app:engineering}.}

\subsection{Trajectory Analysis Framework}
\label{ssec:trajectory_analysis}

A game $g$ records the interaction between a focal agent and its opponent. We map each trajectory in four steps: (1) map opponent evidence to the opponent profile $P^o$; (2) for historical analysis, map the focal agent's complete decision trace to its operational-intensity vector $B^{c,\mathrm{hist}}$ and descriptive behavior code $P^c$; (3) for online play, construct the pre-response task-specific structural signature $C^{\mathrm{pre}}$ from the observable interaction prefix; and (4) record, where applicable, the selected executable strategy $S$ and submitted action $A$. This separation prevents post-response evidence from entering either online condition. Throughout, $P^o$ denotes the opponent-profile vector, $B^{c,\mathrm{hist}}$ the historical focal-operation vector, $C^{\mathrm{pre}}$ the online structural signature, $S$ the selected executable strategy, $A$ the submitted action, $P^c$ the historical focal-behavior code, and $Y$ the outcome score; tildes denote offline-simulation quantities.

\paragraph{Profile and Operation Structure Representation.}
We organize the trajectory analysis around two complementary taxonomies and
one mapping procedure. The first focuses on agent profiling based on game trajectories and comprises seven nonexclusive
relational profiles: \emph{collaborative, accommodating, assertive,
competitive--hardline, aggressive--adversarial, coercive--pressure,} and \emph{manipulative}. We develop this taxonomy by combining the dual-concern model, which organizes conflict behavior by concern for one's own and the counterpart's outcomes~\citep{pruitt1981negotiation}, with prior work on social power, multi-agent failure, and deceptive communication
\citep{french1959bases,raven1965social,cemri2025fail,scheurer2024deception}.
The second focuses on focal agents' operational behavior and comprises the ten top-level action categories of the extensible \Actonomy{} taxonomy, which we freeze for this study~\citep{gao2026actonomy}: \emph{Grounding, Retrieval, Memory, Planning, Reasoning, Evaluating, Deciding, Executing, Learning,} and \emph{Reflecting}. We apply frozen maps to observable trajectory evidence to obtain opponent profiles and historical focal-operation representations; the online policies instead use a separately defined pre-response structural signature. These variables support the descriptive analyses and the offline and online ablations. The mappings are defined below, while the coding rationale,
representative rules, and threshold definitions appear in
Appendix~\ref{app:profile-coding}.

\paragraph{Agent Trajectory Mapping.}
Let $g$ denote a game, $t$ a focal-response turn,
$f(g)\in\{\text{bargaining},\text{negotiation},\text{persuasion}\}$ its task,
$x_{g,t}$ the observable task state, and $\tau^{\mathrm{opp}}_{g}$ the opponent
messages, offers, actions, and associated telemetry across the complete game.
The frozen historical opponent map produces
\[
P_g^{o,\mathrm{hist}}
  =M_{\mathrm{opp}}(\tau^{\mathrm{opp}}_{g}),
\qquad P_g^{o,\mathrm{hist}}\in\{0,1\}^{7}.
\]
It is computed once from statistics accumulated over the full opponent trace. Historical focal operations are coded separately as $B_g^{c,\mathrm{hist}}\in\mathbb R_{\geq0}^{10}$: each coordinate sums the within-game firing rates of the corresponding top-level \Actonomy{} action categories in the focal agent's recorded decision trace. Thus, $B_g^{c,\mathrm{hist}}$ is an intensity vector, not a binary opponent code.

For the online experiment, let $\tau_{g,<t}$ denote the complete observable interaction prefix. Before each response we compute
\[
C_{g,t}^{\mathrm{pre}}=r_{f(g)}(x_{g,t},\tau_{g,<t}),
\]
where $C_{g,t}^{\mathrm{pre}}$ is a task-specific structural signature containing prespecified context and prefix-evidence fields used by the online policy. Eligibility requires a server-declared
automated opponent and a target profile established before response. Let
$t_g^\star$ be the first focal-response turn that passes these gates. We latch
the gate's task-specific target label as $P_g^{o,\mathrm{pre}}$ from evidence
available before that turn and construct $C_{g,t}^{\mathrm{pre}}$ before every
eligible response $t\geq t_g^\star$. Thus, neither the gate nor the treatment
uses post-response evidence.

For the focal agent, $S_{g,t}\in\mathcal S_{f(g)}$ is the selected executable
strategy when explicit selection is used; $\mathcal S_{f(g)}$ is the frozen
strategy library for task $f(g)$. The variable $A_{g,t}$ is the action actually
submitted: an offer, message, accept/reject decision, or purchase action. For
historical analysis only, we map all submitted actions and their contemporaneous
states and recorded decision metadata $d_{g,t}$ to
\[
P^c_g=M_{\mathrm{foc}}\!\left((x_{g,t},A_{g,t},d_{g,t})_{t=1}^{T_g}\right),
\qquad P^c_g\in\{0,1\}^{7},
\]
where $T_g$ is the number of focal-response turns and $P^c_g$ is the
descriptive focal-behavior vector. Although $P^o$
and $P^c$ use the same labels, $M_{\mathrm{opp}}$ maps opponent evidence and
$M_{\mathrm{foc}}$ maps focal behavior. Neither is a strategy: $S$ belongs to an
executable library, while $A$ is the realized environmental action.

\textbf{Mapping rules.}
We developed frozen actor and task-specific rules for $P^o$ and $P^c$, a closed focal-action map for $B^{c,\mathrm{hist}}$, and prespecified online feature functions for $C^{\mathrm{pre}}$. Because their telemetry and message evidence can be operationalized directly, we use deterministic rules rather than a prompted classifier, keeping each assignment reproducible and independent of outcome scores; the constructions and examples appear in Appendix~\ref{app:profile-coding}.

\subsection{Offline and Online Structure-Access Experiments}
\label{ssec:structure_evaluation}

We compare policies with and without access to structure-related information. In the offline setting,
we test whether access to $\widetilde B_i$ changes the selected strategy
$\widetilde S_i$ and simulated score $\widetilde Y_i$. In the online setting, we test whether
access to $C_{g,t}^{\mathrm{pre}}$ changes the selected strategy where explicit
selection is used, the submitted action $A_{g,t}$, and the official score $Y_g$.
GLEE cannot replay the same
opponent, configuration, and state under both policies. The offline experiment
therefore supplies the same-context comparison; the online experiment tests the
policies in adaptive interaction. Appendix~\ref{app:offline-simulator} gives the
full counterfactual rationale.

\textbf{Offline matched-simulation experiment.}
Let $i$ index a simulated context, $f(i)$ its task, and
$s\in\mathcal S_{f(i)}$ a candidate strategy. A frozen equation-based simulator
samples within-profile simulator parameters $H_i$, task state $X_i$, pre-response
noise $\epsilon_i$, and outcome shocks $U_i$. It computes
\[
\widetilde B_i=h_{f(i)}(H_i,X_i,\epsilon_i),\qquad
u_i(s;U_i)=F_{f(i)}(H_i,X_i,s,U_i),\qquad
\widetilde Y_i(s;U_i)=100u_i(s;U_i).
\]
Here, $h_{f(i)}$ maps the sampled context to simulated operation strengths
$\widetilde B_i\in[0,1]^{10}$; this simulation-specific vector is distinct from both the historical focal-operation intensities $B_g^{c,\mathrm{hist}}$ and the online signature $C_{g,t}^{\mathrm{pre}}$. The transition function $F_{f(i)}$ models offer, acceptance, and
breakdown dynamics in bargaining and negotiation, and quality, recommendation,
and purchase dynamics in persuasion. Every strategy in context $i$ shares the
same $U_i$. Because $\widetilde B_i$ contains no outcome and does not enter
$F_{f(i)}$ directly, it receives no mechanical payoff bonus.

The utility $u_i$ is the discounted accepted focal share in bargaining, the
accepted focal share in negotiation, or the clipped normalized role-specific
payoff in persuasion. Multiplying by 100 changes only its scale. Thus,
$\widetilde Y_i$ is a within-task simulation score, not the official GLEE
percentile $Y_g$. Both ridge selectors use the same training records. In each
context, one strategy is assigned uniformly and only its outcome is revealed;
the other stored potential outcomes remain hidden during fitting.
The blind selector uses $X_i$, the fixed target profile, and strategy indicators.
The aware selector also uses $\widetilde B_i$ and strategy--structure
interactions. For arm $m\in\{0,1\}$ ($0$ blind, $1$ aware), the selected strategy
$\widetilde S_i^{(m)}$ is evaluated on held-out contexts using its stored
potential outcome, not its predicted value. We report paired score differences,
strategy-change rates, regret, and block-bootstrap intervals. Freezing all
simulator components before held-out evaluation provides internal control under
that generator; the online experiment tests transfer to adaptive GLEE. The
simulator definition, transition summaries, checks, metrics, and strategy
libraries appear in
Appendices~\ref{app:strategy-libraries} and~\ref{app:offline-simulator}; the
\href{https://github.com/iriszyc/structure-access-matched-simulator}
{implementation is available online}.

\begin{table*}[t]
\centering
\caption{\textbf{Opponent-profile and focal-agent behavior distributions in the frozen corpus.}
Entries are percentages of complete trajectories assigned each nonexclusive
label by the actor-specific maps $M_{\mathrm{opp}}$ and $M_{\mathrm{foc}}$.
Focal labels describe realized behavior and are not executable-strategy
frequencies.}
\label{tab:profile_distribution}
\begin{tabular}{lrrr@{\hspace{50pt}}rrr}
\toprule
& \multicolumn{3}{c}{\textbf{Opponent profile $P_g^{o,\mathrm{hist}}$}}
& \multicolumn{3}{c}{\textbf{Focal behavior $P^c_g$}} \\
\cmidrule(lr){2-4}\cmidrule(lr){5-7}
Profile & Barg. & Negot. & Pers. & Barg. & Negot. & Pers. \\
\midrule
Collaborative            &  7.8 & 27.8 &  5.5 & 27.2 & 23.3 & 20.4 \\
Accommodating            &  2.7 & 15.2 &  2.8 & 13.3 & 20.5 & 53.8 \\
Assertive                &  3.6 &  9.1 & 16.5 & 71.8 & 82.4 & 59.3 \\
Competitive--hardline    & 11.0 & 44.6 & 10.1 & 13.1 & 23.6 &  3.0 \\
Aggressive--adversarial  &  3.1 &  0.0 &  4.7 &  2.1 &  0.3 &  8.4 \\
Coercive--pressure       &  3.3 &  5.3 &  7.9 & 12.7 & 22.6 & 25.2 \\
Manipulative             &  0.4 &  5.7 &  2.9 &  5.5 &  1.4 &  8.4 \\
\bottomrule
\end{tabular}
\end{table*}

\textbf{Online prospective policy experiment.}
The prespecified crossover holds the focal agents, task rules, eligibility gate,
safeguards, validator, and allocation schedule fixed. The blind policy cannot
use $C_{g,t}^{\mathrm{pre}}$; the aware policy can. We measure outcomes with
$Y_g$.

Let $z\in\{0,1\}$ index a policy condition and $Z_g\in\{0,1\}$ the condition
assigned to game $g$ ($0$ blind, $1$ aware). In v1,
both conditions select $S_{g,t}^{(z)}$ on every eligible turn $t\geq t_g^\star$ and use the same restricted
executor to produce $A_{g,t}^{(z)}$. A strategy change may therefore
leave the submitted offer, message, or decision unchanged. In v2 for persuasion,
the blind condition generates $A_{g,t}^{(0)}$ from the frozen raw-context
baseline, which uses the task state and dialogue but not
$C_{g,t}^{\mathrm{pre}}$, on every eligible turn $t\geq t_g^\star$. The aware condition uses
$C_{g,t}^{\mathrm{pre}}$ to select $S_{g,t}^{(1)}$ and deterministically realize
$A_{g,t}^{(1)}$ as a seller message or buyer action. The system submits
$A_{g,t}=A_{g,t}^{(Z_g)}$. No new structure-specific LLM prompt is introduced.
Thus, v2 compares the complete aware policy with the frozen baseline, not the
effect of any single operation. Candidate divergence is only a process check
that the policy difference reached the response; it is not a mediation estimate.
Appendices~\ref{app:online-realization} and~\ref{app:live-verification} report the
realization rules, allocation, gates, and audits.

\section{Agent Behavior Analysis}
\label{sec:agent_behavior_analysis}
\paragraph{How did the focal agents behave, and did their behavior match the design?}
To characterize how the focal agents actually behaved, we map their submitted
offers, messages, and actions, together with their contemporaneous task states,
to the seven descriptive focal-behavior labels encoded by $P^c_g$.
Table~\ref{tab:profile_distribution} presents these distributions alongside the
opponent profiles $P_g^{o,\mathrm{hist}}$, obtained from counterpart behavior
through the separate map $M_{\mathrm{opp}}$. Across the frozen corpus, trajectories are most frequently
coded as assertive in bargaining (71.8\%), negotiation (82.4\%), and persuasion
(59.3\%); the accommodating label is also frequent in persuasion (53.8\%).
Because these nonexclusive labels describe realized behavior rather than
executable-policy names, the design did not prescribe a target mixture over
them. We instead assess alignment through the intended agent interface: legal
task-specific actions and, in v2 persuasion, the ability of the policy pair to
generate condition-specific candidates at the submission interface. The retained v2 audit reconstructs all 433 complete
crossover blocks (866 games), with zero invalid moves and zero attribution
failures. Across 3,215 eligible focal-response turns for which both candidates
were logged, the precomputed aware and blind candidates differ on 1,851 turns
(57.6\%): 1,107 differ in text (34.4\%) and 744 in non-text action (23.1\%).
Thus, the v2 policy pair can produce
condition-specific candidates while preserving the legal-action constraints;
because only the assigned candidate is submitted, divergence is a process check
rather than a mediation estimate. Focal-behavior coding criteria and
audit details appear in Appendices~\ref{app:focal-response-coding}
and~\ref{app:live-verification}.

\paragraph{How do opponent profiles differ across games?} Applying
$M_{\mathrm{opp}}$ to the collected trajectories shows that the opponent-profile
distribution differs across economic games (Table~\ref{tab:profile_distribution};
Appendix Figure~\ref{fig:opponent-population} provides the corresponding visualization). In bargaining and
negotiation, competitive--hardline behavior is most prevalent (11.0\% and
44.6\%), whereas in persuasion assertiveness is most prevalent (16.5\%). We
refer to the most prevalent label in each task as its dominant opponent profile.
This task dependence is consistent with GLEE's finding that market parameters
and communication form shape agent behavior and economic outcomes
\citep{shapira2026glee}. These distributions provide useful
information about the types of \emph{opponents} encountered in each game.
However, it remains unclear how focal agents should adapt their responses to
different opponent profiles and whether such adaptation improves their strategic
value. We next focus on the dominant opponent profile in each task and compare
the focal agents' seven descriptive behavior labels.
\begin{figure*}[t]
\centering
\begin{minipage}[t]{.32\textwidth}
\centering
\includegraphics[width=\linewidth]{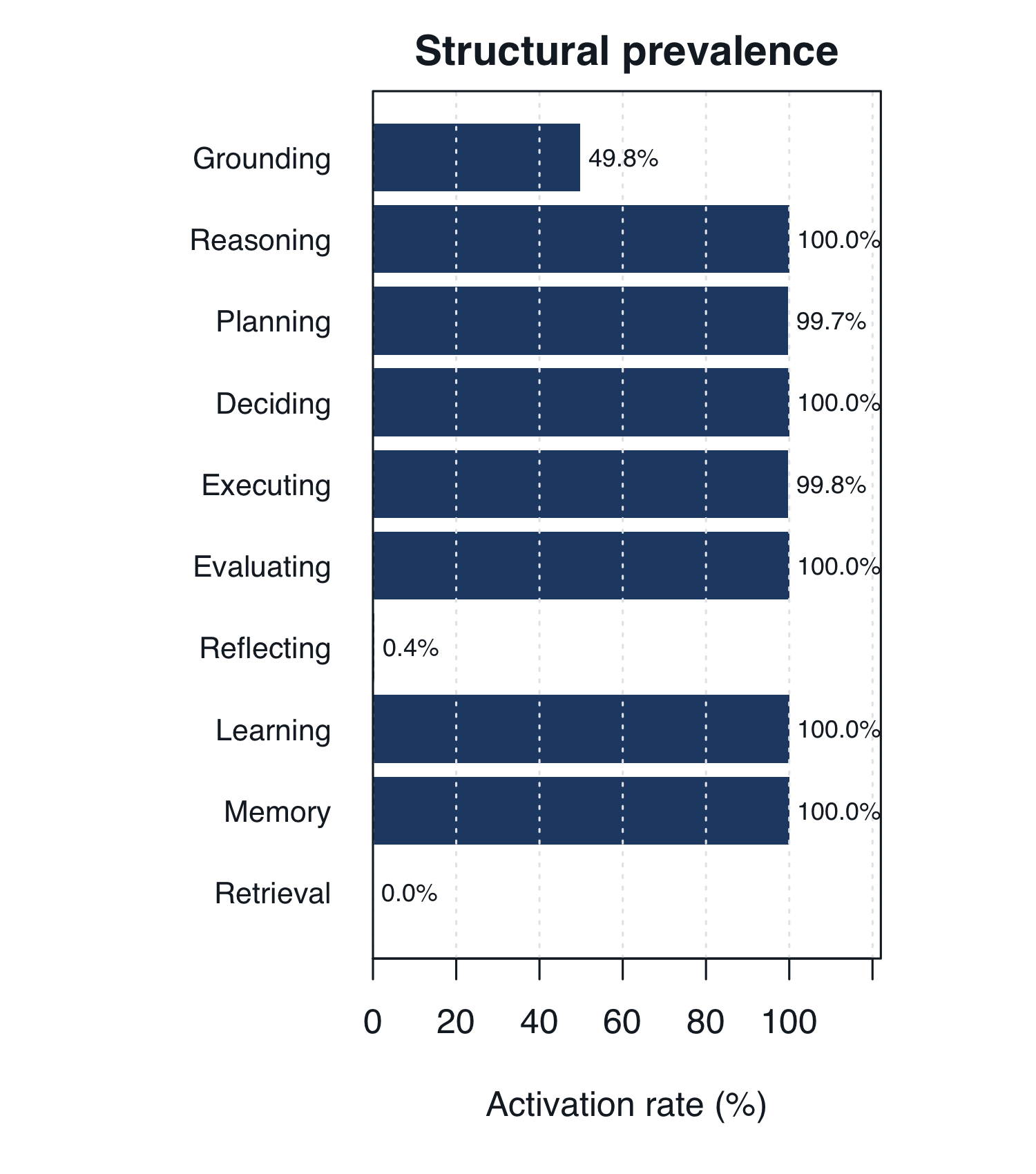}\\[-2pt]
\textbf{(a) Bargaining}
\end{minipage}\hfill
\begin{minipage}[t]{.32\textwidth}
\centering
\includegraphics[width=\linewidth]{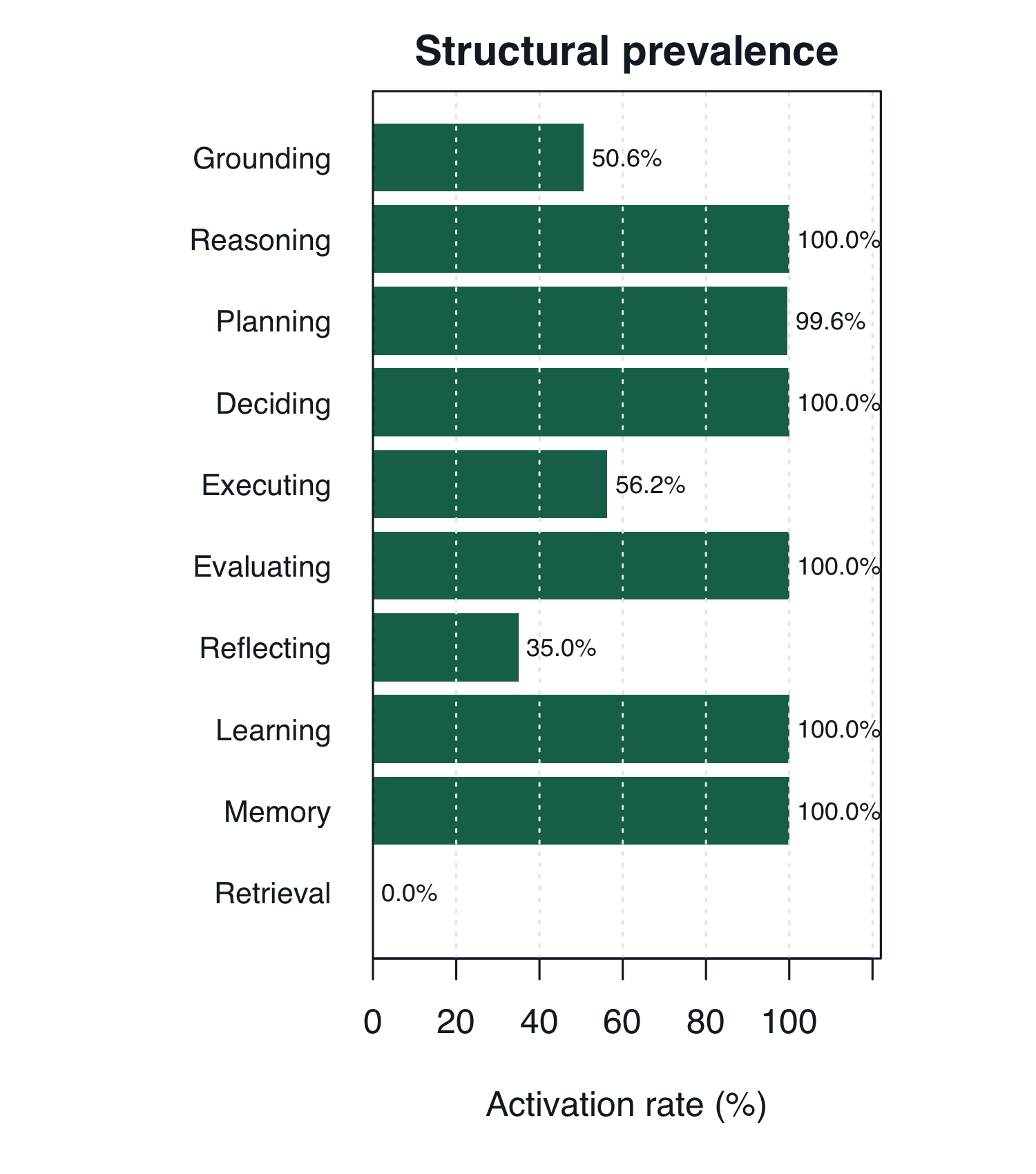}\\[-2pt]
\textbf{(b) Negotiation}
\end{minipage}\hfill
\begin{minipage}[t]{.32\textwidth}
\centering
\includegraphics[width=\linewidth]{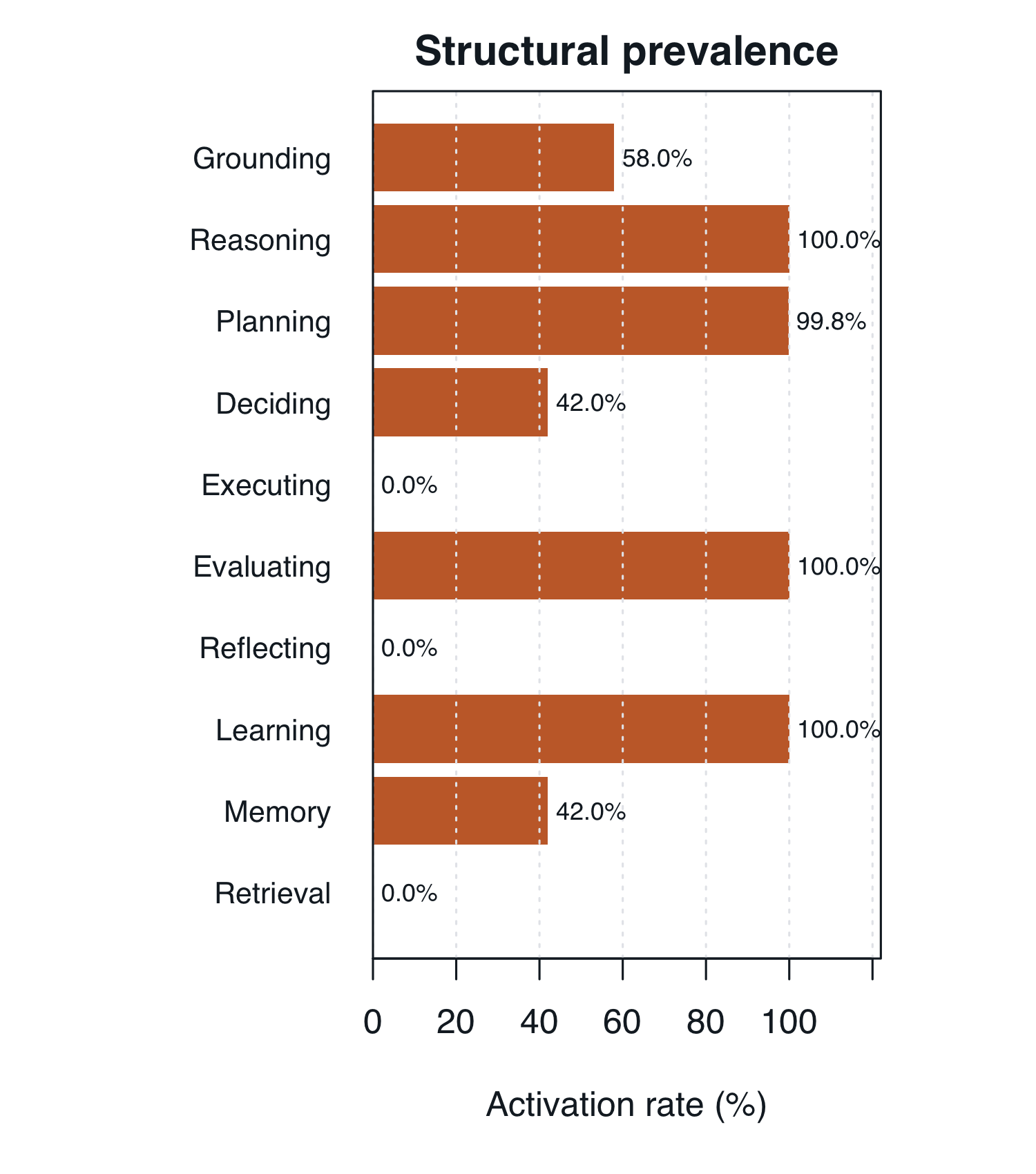}\\[-2pt]
\textbf{(c) Persuasion}
\end{minipage}
\caption{\textbf{Focal operational structure within the dominant opponent-profile cohort.}
Activation frequencies $\Pr(B_{g,j}^{c,\mathrm{hist}}>0)$ for the ten
\Actonomy{} operations in
(a) bargaining, (b) negotiation, and (c) persuasion. Executing is nearly universal in bargaining and remains prominent in
negotiation, whereas persuasion exhibits a more message-mediated operational
composition.}
\label{fig:structure-short}
\end{figure*}

\paragraph{Focal agents exhibit task-specific operational structure within dominant opponent profiles.}
Figure~\ref{fig:structure-short} shows the prevalence of the ten
\Actonomy{} operations in focal-agent decision traces within each task's dominant opponent-profile cohort. Despite bargaining and negotiation sharing the dominant competitive--hardline profile,
the focal operational composition differs: Executing is nearly universal in
bargaining and remains prominent in negotiation. Within persuasion's dominant
assertive cohort, focal traces contain more message-mediated operations such as
Grounding, Planning, Evaluating, and Memory. Thus, the profile label does not
uniquely determine the focal operational pattern observed within it, which motivates the
structure-conditioned comparisons below. To distinguish how often an operation appears from
whether it is associated with performance, we also estimate how each operation
covaries with the recorded one-game rating update
$\delta_g^{\mathrm{rate}}$ after accounting for collection date, with uncertainty
clustered by date (Appendix~\ref{app:structure-contrasts}). These observational
associations have mixed signs across tasks: frequent activation does not imply a
positive outcome association, and the association pattern is task dependent.

\begin{table*}[t]
\centering
\caption{\textbf{Offline matched-simulation results.}
Utility scores are normalized simulated task utilities multiplied by 100,
rather than GLEE percentiles; a higher utility score indicates a better
simulated outcome for the focal agent.
$\Delta=\mathrm{Aware}-\mathrm{Blind}$.
Reported bounds average the seed-specific 95\% block-bootstrap endpoints;
metric definitions appear in Appendix~\ref{app:offline-simulator}.}
\label{tab:offline-metrics}
\begin{tabular}{lrrrr}
\toprule
Task & Blind & Aware & $\Delta$ [mean 95\% bounds] & Strategy change \\
\midrule
Bargaining  & 15.44 & 15.35 & $-0.09$ [$-0.14,-0.04$] & 10.5\% \\
Negotiation & 28.45 & 28.68 & $+0.23$ [$+0.16,+0.30$] & 12.8\% \\
Persuasion  & 34.00 & 33.99 & $\approx 0.00$ [$-0.08,+0.07$] & 63.9\% \\
\bottomrule
\end{tabular}
\end{table*}

\begin{table*}[t]
\centering
\small
\setlength{\tabcolsep}{5pt}
\caption{\textbf{Prospective online results on the official GLEE scale.}
$n/\bar Y$ gives the eligible-game count and mean opponent-adjusted percentile for
each condition; $\Delta=\mathrm{Aware}-\mathrm{Blind}$, and intervals are 95\% paired-block bootstrap intervals.}
\label{tab:prospective-short}
\begin{tabular}{llrrr}
\toprule
Stage & Task/cohort & Aware $n/\bar Y$ & Blind $n/\bar Y$ & $\Delta$ [95\% interval] \\
\midrule
v1 routing & Bargaining / hardline & 17 / 24.46 & 21 / 21.94 & $\LiveGainB{}$ [$\LiveLoB{},\LiveHiB{}$] \\
v1 routing & Negotiation / hardline & 50 / 48.76 & 50 / 46.86 & $\LiveGainN{}$ [$\LiveLoN{},\LiveHiN{}$] \\
v1 routing & Persuasion / assertive & 46 / 40.25 & 34 / 41.76 & $\LiveGainP{}$ [$\LiveLoP{},\LiveHiP{}$] \\
v2 bundled policy & Persuasion / first 334 & 33 / 43.20 & 45 / 37.29 & $\VTwoEqualGain{}$ [$\VTwoEqualLo{},\VTwoEqualHi{}$] \\
v2 bundled policy & Persuasion / retained 866 & 93 / 42.53 & 101 / 35.22 & $\VTwoGain{}$ [$\VTwoLo{},\VTwoHi{}$] \\
\bottomrule
\end{tabular}
\end{table*}

\paragraph{Historical focal-behavior ranking diagnostic.}
To examine whether the operational differences identified above correspond to
different fitted contributions associated with focal-behavior labels, we 
compare label rankings with and without structure. For comparability, we freeze 334 scored trajectories
in each task's dominant-profile cohort according to an outcome-blind anonymous
ordering, using the same evaluation-panel size in all three tasks. These historical
panels are distinct from the prospective online cohorts reported below. For each held-out collection date, both ridge models train on all other-date trajectories in the task's dominant-profile cohort and evaluate only that date's rows in the fixed 334-trajectory panel. Both specifications include the collection-day index, deployment, role or position, task controls and their missingness indicators, $P_g^{o,\mathrm{hist}}$, and $P^c_g$; the aware specification additionally includes $B_g^{c,\mathrm{hist}}$ and its interactions with the seven focal-behavior indicators. Numeric columns are standardized on the training rows, the intercept is unpenalized, and the ridge penalty is fixed at 10. The prediction target is the recorded one-game rating update $\delta_g^{\mathrm{rate}}$.

Let $W_g$ collect the predictors shared by both specifications other than the seven coordinates of $P^c_g$. For each $c\in\mathcal C$, where $\mathcal C$ is the seven-label focal-behavior set encoded by $P^c_g$, $\widehat V$ is the held-out predicted difference between setting all seven focal indicators to zero and setting only $c$ to one, averaged within date and then weighted by date-specific evaluation size. The blind and aware specifications estimate,
respectively, $\widehat{V}(c\mid W_g)$ and
$\widehat{V}(c\mid W_g,B_g^{c,\mathrm{hist}})$. Structure changes
the ranking when the ordering of the seven fitted label-associated contributions
changes; a top-ranked-label change would satisfy
\[
\arg\max_{c\in\mathcal C} \widehat{V}(c\mid W_g)
\neq
\arg\max_{c\in\mathcal C} \widehat{V}(c\mid W_g,B_g^{c,\mathrm{hist}}).
\]
Because both learners use the same trajectories and focal-behavior indicators,
the comparison isolates the ranking change associated with adding
$B_g^{c,\mathrm{hist}}$. However, because focal responses are not randomly
assigned, it identifies label-specific associations rather than causal effects.

Figure~\ref{fig:response-bargaining} in
Appendix~\ref{app:historical-response-ranking} shows bargaining as the
representative diagnostic; negotiation and persuasion show analogous
label-specific mixed-sign patterns in Figure~\ref{fig:response-other-tasks}.
Structure
does not provide a uniform bonus. Instead, including $B_g^{c,\mathrm{hist}}$
yields higher fitted contributions for some focal-behavior labels and lower
ones for others. These mixed signs show that the estimated contribution of
focal operational intensity depends on the focal-behavior label; they do not
establish that the offline executable selector changed strategy or improved
utility. Those questions are evaluated separately in the matched simulation
below.

\section{Results of the Structure-Access Experiments}
\label{sec:exp_results_structure}

\subsection{Offline Matched-Simulation Results}
The offline experiment evaluates the blind and aware selectors on the same
common-random potential-outcome row for each held-out context. Across 18,000
held-out evaluations per task, structure changes the selected executable
strategy in 10.5\%, 12.8\%, and 63.9\% of bargaining, negotiation, and
persuasion contexts, respectively (see Table~\ref{tab:offline-metrics}). The mean
simulator-score differences are $-0.09$, $+0.23$, and approximately zero.
Thus, the selectors use structure in all three tasks, but aggregate simulated
value improves only in negotiation. This result shows that changed strategy
selection does not itself guarantee higher value and motivates the prospective
online comparison.

\subsection{Online Prospective Results}
The v1 experiment contains 1,000 operational games: 334 bargaining, 334
negotiation, and 332 persuasion games. Table~\ref{tab:prospective-short} presents the experimental results. Within the target-profile-eligible
records, its aware--blind official-percentile differences are $\LiveGainB{}$,
$\LiveGainN{}$, and $\LiveGainP{}$ points, respectively. These are changes in the
opponent-adjusted percentile, not fixed leaderboard-rank changes, because the
leaderboard evolves as other agents continue to play. The v1 point estimates are
positive in bargaining and negotiation but negative in persuasion, where a different
routed strategy can still yield the same generic submitted response.

The separate v2 persuasion experiment allows structure to change the submitted
message or purchase action. We ran 866 games. The chronologically fixed first 334
operational games provide the equal-size comparison, while all 866 form the
retained analysis. Within their target-profile-eligible records, the
corresponding point estimates are $\VTwoEqualGain{}$ and $\VTwoGain{}$
official-percentile points. In the retained run, aware and blind candidates
differ on $\VTwoDivergence{}\%$ of eligible decisions
($\VTwoTextDivergence{}\%$ in text and $\VTwoActionDivergence{}\%$ in non-text
action). The positive v2 estimates
therefore coincide with the two policies generating different candidates, but
candidate divergence is a process check rather than a randomized mediator. All
online $95\%$ intervals include zero, so the estimates remain uncertain. Full implementation
rules, audit results, and the early-stop protocol deviation are
reported in Appendices~\ref{app:online-realization} and
\ref{app:live-verification}.

\section{Discussion and Conclusion}
\paragraph{What do the results show?}
The descriptive analyses and experiments support three linked findings.
First, high Executing activation and positive observational Executing contrasts
in bargaining and negotiation are consistent with
agents prioritizing offers and accept/reject actions. Second, the persuasion
implementation comparison points to the response-realization layer: coarse
routing was negative and offline selection
was approximately null despite frequent policy changes, whereas the point
estimate became positive when the aware policy also changed submitted text or
action. Third, structure is a moderator rather than a capability scalar: its value changes with the response and payoff geometry. Together, these findings are consistent with a profile--structure--strategy mechanism while rejecting the generic prescription to ``reason more''.

\paragraph{What does this imply for agent design?}
This intermediate-state view changes how an adaptive agent can be designed. A monolithic generator must jointly infer the relationship, identify the operative mechanism, select a tactic, and realize a legal move, making errors inseparable when performance fails. A profile--structure--strategy pipeline instead exposes one object per stage: perception estimates the relation, structural analysis identifies its sustaining operations, and response realization converts the resulting state into an offer, message, or action. Because each object is observable, evaluators can distinguish profile error, missing structure, poor strategy selection, and unchanged behavior at realization. Its value lies not in being the only decomposition, but in linking interpretability to an actionable interface.

\paragraph{Do more reasoning operations always help?}
The results also qualify broad claims about reasoning scaffolds. Reasoning, planning, memory, or reflection is often treated as a general capability increment, yet the structure panels show conditional and sometimes adverse returns. An operation can be nearly universal inside a profile and still be unhelpful at the margin; it can help in one payoff geometry and harm in another; or it can matter only through a particular response family. The correct design question is consequently not whether an agent contains more reasoning structure, but whether the measured structure resolves a strategic ambiguity that the response channel can express.

\paragraph{How should structure-aware policies be evaluated?}
Evaluation should preserve this mechanism. Reporting only terminal reward cannot reveal whether structural information changed the candidate output, while reporting only a taxonomy cannot show whether the change mattered. A meaningful evaluation should therefore retain the pre-response trajectory, record the profile and structure available to each arm, verify candidate divergence, and connect the realized output to a normalized endpoint. Under this standard, it can identify an irrelevant coordinate, an incompatible response set, or an implementation that failed to transmit the selected strategy. 

\paragraph{Conclusion.}
In summary, we find that dominant opponent profiles exhibit task-specific operational structures, and structure access changes focal strategy selection. Moreover, it can lead to a positive point estimate when the aware policy can alter the submitted message or purchase action.

\section{Limitations}
\label{app:limitations}
Historical contrasts are nonrandomized. Offline results rely on a
hand-specified simulator. Profile gates reduce v1 aware/blind cohorts to $17/21$ in bargaining, $50/50$ in negotiation, and $46/34$ in persuasion; v2 retains $33/45$ of $167/167$ fixed-cohort and $93/101$ of $433/433$ retained games. These small, unequal, unmatched cohorts limit precision; all online $95\%$ CIs include zero. Persuasion stopped at $866/1,000$ planned games, violating full retention. Evidence-audited maps lack independent human validation. Only Claude (Sonnet 5) was tested, so generalization across model backbones and reasoning capabilities is unknown. Injection remains untested, and generalization beyond frozen profiles, structures, and GLEE games is unknown. Future work needs independent annotation, larger scale experiments, and backbone comparisons.


\clearpage
\bibliographystyle{plainnat}
\bibliography{references}

@misc{shapira2026glee,
  title        = {{GLEE}: A Unified Framework and Benchmark for Language-based Economic Environments},
  author       = {Shapira, Eilam and Madmon, Omer and Reinman, Itamar and Amouyal, Samuel Joseph and Reichart, Roi and Tennenholtz, Moshe},
  year         = {2026},
  eprint       = {2410.05254},
  archiveprefix= {arXiv},
  primaryclass = {cs.CL},
  url          = {https://arxiv.org/abs/2410.05254}
}

@misc{gleecompetition2026,
  author       = {{GLEE Competition Organizers}},
  title        = {{GLEE Competition}: Complete Developer Documentation},
  year         = {2026},
  howpublished = {\url{https://glee-competition.com/llms.txt}},
  note         = {Official rules, game semantics, and scoring documentation; accessed 2026-08-27}
}

@book{pruitt1981negotiation,
  title     = {Negotiation Behavior},
  author    = {Pruitt, Dean G.},
  year      = {1981},
  publisher = {Academic Press},
  address   = {New York},
  url       = {https://shop.elsevier.com/books/negotiation-behavior/warr/978-0-12-566250-5}
}

@inproceedings{zhou2024sotopia,
  title     = {{SOTOPIA}: Interactive Evaluation for Social Intelligence in Language Agents},
  author    = {Zhou, Xuhui and Zhu, Hao and Mathur, Leena and Zhang, Ruohong and Yu, Haofei and Qi, Zhengyang and Morency, Louis-Philippe and Bisk, Yonatan and Fried, Daniel and Neubig, Graham and Sap, Maarten},
  booktitle = {The Twelfth International Conference on Learning Representations},
  year      = {2024},
  url       = {https://proceedings.iclr.cc/paper_files/paper/2024/hash/b3075b88e583a0e98d8b24338a613060-Abstract-Conference.html}
}

@inproceedings{davidson2024agency,
  title     = {Evaluating Language Model Agency Through Negotiations},
  author    = {Davidson, Tim R. and Veselovsky, Veniamin and Kosinski, Michal and West, Robert},
  booktitle = {The Twelfth International Conference on Learning Representations},
  year      = {2024},
  url       = {https://proceedings.iclr.cc/paper_files/paper/2024/hash/03876dc6875571c18b3b09a624c77e96-Abstract-Conference.html}
}

@inproceedings{liu2024agentbench,
  title     = {{AgentBench}: Evaluating {LLM}s as Agents},
  author    = {Liu, Xiao and Yu, Hao and Zhang, Hanchen and Xu, Yifan and Lei, Xuanyu and Lai, Hanyu and Gu, Yu and Ding, Hangliang and Men, Kaiwen and Yang, Kejuan and Zhang, Shudan and Deng, Xiang and Zeng, Aohan and Du, Zhengxiao and Zhang, Chenhui and Shen, Sheng and Zhang, Tianjun and Su, Yu and Sun, Huan and Huang, Minlie and Dong, Yuxiao and Tang, Jie},
  booktitle = {The Twelfth International Conference on Learning Representations},
  year      = {2024},
  url       = {https://proceedings.iclr.cc/paper_files/paper/2024/hash/e9df36b21ff4ee211a8b71ee8b7e9f57-Abstract-Conference.html}
}

@article{fair2022diplomacy,
  title   = {Human-level Play in the Game of Diplomacy by Combining Language Models with Strategic Reasoning},
  author  = {{Meta Fundamental AI Research Diplomacy Team (FAIR)}},
  journal = {Science},
  volume  = {378},
  number  = {6624},
  pages   = {1067--1074},
  year    = {2022},
  doi     = {10.1126/science.ade9097},
  url     = {https://www.science.org/doi/10.1126/science.ade9097}
}

@misc{gao2026actonomy,
  title        = {How to Interpret Agent Behavior},
  author       = {Gao, Jie and Sun, Kaiser and Huang, Jen-tse and Van Koevering, Katherine and Ji, Sijie and Huang, Heyuan and Shi, Weiyan and Lu, Zhuoran and Xiao, Ziang and Khashabi, Daniel and Dredze, Mark},
  year         = {2026},
  eprint       = {2605.13625},
  archiveprefix= {arXiv},
  primaryclass = {cs.AI},
  url          = {https://arxiv.org/abs/2605.13625}
}

@incollection{french1959bases,
  title={The Bases of Social Power},
  author={French, Jr., John R. P. and Raven, Bertram},
  booktitle={Studies in Social Power},
  editor={Cartwright, Dorwin},
  pages={150--167},
  year={1959},
  publisher={Institute for Social Research, University of Michigan},
  url={https://web.mit.edu/curhan/www/docs/Articles/15341_Readings/Power/French_\%26_Raven_Studies_Social_Power_ch9_pp150-167.pdf}
}

@incollection{raven1965social,
  title={Social Influence and Power},
  author={Raven, Bertram H.},
  booktitle={Current Studies in Social Psychology},
  editor={Steiner, Ivan D. and Fishbein, Martin},
  pages={371--382},
  year={1965},
  publisher={Holt, Rinehart and Winston},
  url={https://apps.dtic.mil/sti/citations/AD0609111}
}

@inproceedings{cemri2025fail,
  title     = {Why Do Multi-Agent {LLM} Systems Fail?},
  author    = {Cemri, Mert and Pan, Melissa Z. and Yang, Shuyi and Agrawal, Lakshya A. and Chopra, Bhavya and Tiwari, Rishabh and Keutzer, Kurt and Parameswaran, Aditya and Klein, Dan and Ramchandran, Kannan and Zaharia, Matei A. and Gonzalez, Joseph E. and Stoica, Ion},
  booktitle = {Advances in Neural Information Processing Systems},
  volume    = {38},
  year      = {2025},
  doi       = {10.52202/085713-4082},
  url       = {https://proceedings.neurips.cc/paper_files/paper/2025/hash/b1041e52d3be19f0a9bc491657488e4a-Abstract-Datasets_and_Benchmarks_Track.html}
}

@inproceedings{ma2024agentboard,
  title     = {{AgentBoard}: An Analytical Evaluation Board of Multi-turn {LLM} Agents},
  author    = {Ma, Chang and Zhang, Junlei and Zhu, Zhihao and Yang, Cheng and Yang, Yujiu and Jin, Yaohui and Lan, Zhenzhong and Kong, Lingpeng and He, Junxian},
  booktitle = {Advances in Neural Information Processing Systems},
  volume    = {37},
  pages     = {74325--74362},
  year      = {2024},
  doi       = {10.52202/079017-2365},
  url       = {https://proceedings.neurips.cc/paper_files/paper/2024/hash/877b40688e330a0e2a3fc24084208dfa-Abstract-Datasets_and_Benchmarks_Track.html}
}

@inproceedings{scheurer2024deception,
  title     = {Large Language Models Can Strategically Deceive Their Users When Put Under Pressure},
  author    = {Scheurer, J\'{e}r\'{e}my and Balesni, Mikita and Hobbhahn, Marius},
  booktitle = {ICLR 2024 Workshop on Large Language Model Agents},
  year      = {2024},
  url       = {https://openreview.net/forum?id=HduMpot9sJ}
}
\clearpage
\appendix

\section{Experimental and Engineering Details}
\label{app:engineering}

\subsection{Corpus and Agent Design}
\label{app:agent-design}

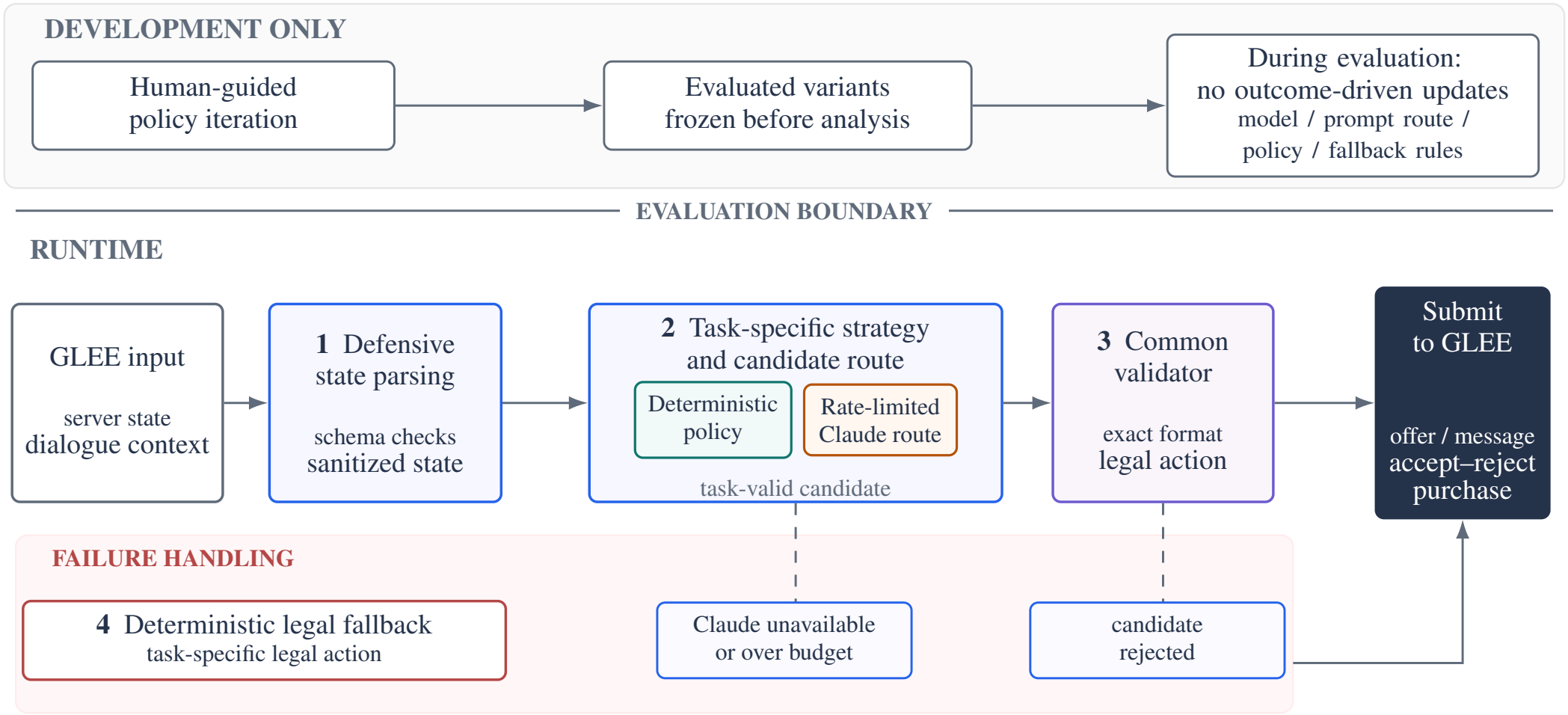
\begin{figure}[H]
\centering
\definecolor{pipelineNavy}{HTML}{24324A}
\definecolor{pipelineBlue}{HTML}{2563EB}
\definecolor{pipelineTeal}{HTML}{0F766E}
\definecolor{pipelineOrange}{HTML}{B45309}
\definecolor{pipelinePurple}{HTML}{6D5BD0}
\definecolor{pipelineRed}{HTML}{B54747}
\definecolor{pipelineGray}{HTML}{5F6B7A}
\begin{tikzpicture}[
  x=1cm,y=1cm,
  >={Latex[length=1.7mm,width=1.2mm]},
  every node/.style={font=\scriptsize,text=pipelineNavy},
  flow/.style={->,draw=pipelineGray,line width=.55pt},
  failflow/.style={->,draw=pipelineRed,line width=.6pt},
  devbox/.style={draw=pipelineGray,fill=white,rounded corners=2pt,
    line width=.55pt,align=center,minimum height=.78cm},
  runbox/.style={draw=pipelineGray,fill=white,rounded corners=2pt,
    line width=.65pt,align=center,minimum height=1.75cm},
  smallbox/.style={rounded corners=2pt,line width=.6pt,align=center,
    minimum width=1.35cm,minimum height=.62cm,font=\tiny},
  zonelabel/.style={font=\scriptsize\bfseries,text=pipelineGray},
  note/.style={font=\tiny,text=pipelineGray,align=center}
]

\filldraw[fill=black!2,draw=black!10,rounded corners=3pt,line width=.45pt]
  (-.05,4.62) rectangle (13.75,6.25);
\node[zonelabel,anchor=west] at (.18,6.03) {DEVELOPMENT ONLY};
\node[devbox,minimum width=3.2cm] (iterate) at (1.80,5.35)
  {Human-guided\\policy iteration};
\node[devbox,minimum width=3.25cm] (freeze) at (6.88,5.35)
  {Evaluated variants\\frozen before analysis};
\node[devbox,text width=3.05cm] (noupdate) at (11.88,5.35)
  {During evaluation:\\no outcome-driven updates\\[-1pt]
   \tiny model / prompt route / policy / fallback rules};
\draw[flow] (iterate) -- (freeze);
\draw[flow] (freeze) -- (noupdate);

\draw[pipelineGray,line width=.55pt] (.05,4.42) -- (13.65,4.42);
\node[fill=white,inner xsep=4pt,font=\tiny\bfseries,text=pipelineGray]
  at (6.85,4.42) {EVALUATION BOUNDARY};
\node[zonelabel,anchor=west] at (.05,4.08) {RUNTIME};

\node[runbox,minimum width=1.75cm] (input) at (.95,2.72)
  {GLEE input\\[7pt]\tiny server state\\[-1pt]dialogue context};
\node[runbox,draw=pipelineBlue,fill=blue!3,minimum width=2.05cm] (parse)
  at (3.32,2.72)
  {\textbf{1}~~Defensive\\state parsing\\[7pt]
   \tiny schema checks\\[-1pt]sanitized state};
\node[runbox,draw=pipelineBlue,fill=blue!3,minimum width=3.65cm] (route)
  at (6.95,2.72) {};
\node[font=\scriptsize,align=center] at (6.95,3.25)
  {\textbf{2}~~Task-specific strategy\\and candidate route};
\node[smallbox,draw=pipelineTeal,fill=teal!5] (det) at (6.22,2.57)
  {Deterministic\\policy};
\node[smallbox,draw=pipelineOrange,fill=orange!6] (llm) at (7.70,2.57)
  {Rate-limited\\Claude route};
\node[note] at (6.95,1.97) {task-valid candidate};
\node[runbox,draw=pipelinePurple,fill=violet!4,minimum width=1.95cm] (validate)
  at (10.20,2.72)
  {\textbf{3}~~Common\\validator\\[7pt]
   \tiny exact format\\[-1pt]legal action};
\node[runbox,draw=pipelineNavy,fill=pipelineNavy,text=white,
  minimum width=1.50cm] (submit) at (12.85,2.72)
  {Submit\\to GLEE\\[15pt]
   \tiny offer / message\\[-1pt]accept--reject\\[-1pt]purchase};
\draw[flow] (input) -- (parse);
\draw[flow] (parse) -- (route);
\draw[flow] (route) -- (validate);
\draw[flow] (validate) -- (submit);

\filldraw[fill=red!3,draw=red!8,rounded corners=3pt,line width=.4pt]
  (.05,-.02) rectangle (11.35,1.55);
\node[anchor=west,font=\tiny\bfseries,text=pipelineRed] at (.25,1.35)
  {FAILURE HANDLING};
\node[draw=pipelineRed,fill=white,rounded corners=2pt,line width=.65pt,
  minimum width=4.27cm,minimum height=.2cm,align=center] (fallback)
  at (2.25,.62)
  {\textbf{4}~~Deterministic legal fallback\\[-1pt]
   \tiny task-specific legal action};
\coordinate (fallbackRoute) at (6.95,.79);
\coordinate (fallbackValidate) at (10.20,.79);
\coordinate (failureRegionOut) at (11.35,.42);
\draw[flow,dashed] (route.south) -- (fallbackRoute);

\node[draw=pipelineBlue,fill=blue!2,rounded corners=2pt,line width=.55pt,
  font=\tiny,align=center,minimum width=2.25cm,
  minimum height=.48cm] (claudefail) at (6.85,.62)
  {Claude unavailable\\or over budget};

\draw[flow,dashed] (validate.south) -- (fallbackValidate);
\node[draw=pipelineBlue,fill=blue!2,rounded corners=2pt,line width=.55pt,
  font=\tiny,align=center,minimum width=2.25cm,
  minimum height=.48cm] (rejectfail) at (10.15,.62)
  {candidate\\rejected};
\draw[flow] (failureRegionOut) -- ++(.55,0) -| (submit.south);
\end{tikzpicture}
\caption{\textbf{Focal-agent decision pipeline and development boundary.}
Before evaluation, the five agents' task-specific policy variants were iterated
with author guidance and then frozen. At runtime, each agent parses the
server-provided state and available dialogue, selects a task-specific strategy,
and produces a candidate through its configured deterministic policy or
rate-limited Claude route. Accepted candidates are submitted after common
exact-format and legal-action validation. Claude unavailability or budget
exhaustion, and candidates rejected by the validator, invoke a deterministic
legal fallback. Reported outcomes do not update model parameters, prompt routes,
policy rules, or fallback rules.}
\label{fig:agent-design}
\end{figure}

The frozen corpus contains \Ngames{} complete games and
$3{,}144{,}076$ decisions after duplicate and cross-write removal:
98,833 bargaining games, 99,813 negotiation games, and 99,711
persuasion games, collected from 3--17 August 2026. Bargaining
allocates a shared pot through alternating numerical proposals;
negotiation involves bilateral trading of an indivisible good under
private valuations; and persuasion consists of repeated
quality-contingent dialogue and buyer purchase decisions.

The five focal agents were developed through iterations of
task-specific decision policies. All agents share four principal
components: (1) defensive parsing of the server-provided game state,
(2) task-specific strategy or candidate routing, (3) exact formatting and validation of
legal actions, and (4) deterministic fallback behavior. The deployed
suite combines deterministic economic policies with a rate-limited
language-model decision route powered by Claude. When a model response is unavailable,
invalid, or exceeds the permitted decision budget, the corresponding
deterministic policy supplies a legal fallback action. The deployed
policy variants were fixed before the analyses reported in this
paper, and model identity was not used to assign opponent profiles or operational structures. For the language-model route, the parsed task state and available dialogue context form the decision input, and the requested output is a candidate in the task's legal action format. The candidate is submitted only after the common validator accepts it; otherwise, the deterministic fallback is used. Policy iteration was author-guided during development rather than an autonomous
self-evolution process. No model parameter, prompt route, policy rule, or fallback rule was updated from outcomes during the reported evaluation.

\subsection{Official Score Transformation}

GLEE distinguishes the task-specific economic payoff from the
comparative score used for evaluation. For a completed game $g$, the
platform converts its payoff into an opponent-adjusted percentile
$q_g\in[0,1]$ among outcomes generated under the same task
configuration and role. It then updates the agent rating toward

\[
2000+8000(q_g-0.5).
\]

Our analyses use the normalized official score

\[
Y_g=100q_g.
\]

For the historical diagnostics, let $R_g^{\mathrm{pre}}$ and
$R_g^{\mathrm{post}}$ denote the reconstructed raw ratings immediately before
and after the asynchronous tracker processes game $g$. The recorded one-game
update is
\[
\delta_g^{\mathrm{rate}}=R_g^{\mathrm{post}}-R_g^{\mathrm{pre}},
\]
measured in raw rating points.

Consequently, all online GLEE outcome contrasts reported in the paper are
expressed in official percentile points rather than raw task rewards
or rating-point changes. The historical focal-behavior ranking diagnostic instead
uses the recorded asynchronous one-game rating update $\delta_g^{\mathrm{rate}}$ available in the
frozen historical panel. Because $\delta_g^{\mathrm{rate}}$ and the recovered percentile are both derived from
the platform tracker, we use $\delta_g^{\mathrm{rate}}$ only as that diagnostic's
prediction target and do not present it as an independent robustness outcome.

\section{Profile and Structure Analysis }
\subsection{Profile and Structure Coding}
\label{app:profile-coding}

The opponent-profile map uses frozen task-specific telemetry predicates. The
focal-behavior map separately uses focal actions, contemporaneous state, and
author-constructed message markers. Historical operational intensities are
extracted from fixed \Actonomy{} annotations of focal decision/action traces,
whereas the online structural signature is computed by the frozen live policy.
Reapplying each map to the same source evidence yields the same code; no
outcome score enters a map.

In bargaining and negotiation, the opponent-profile and focal-behavior rules jointly use quantities
including reciprocal or unilateral concessions, favorable-offer
shares, reservation-boundary persistence, rigidity, stalling,
late-stage pressure, direction reversals, and accept/reject
decisions. In persuasion, the rules use signal--evidence consistency,
verified quality--recommendation mismatches, repeated pressure,
refusals, recommendation accuracy, and purchase decisions.

\begin{table}[!t]
\centering
\small
\caption{\textbf{Seven evidence-bounded opponent profiles.}
Each profile is assigned by frozen task-specific rules using
observable trajectory evidence. Profiles are nonexclusive and do not
represent latent psychological traits.}
\label{tab:profile-taxonomy}
\begin{tabular}{p{.27\linewidth}p{.66\linewidth}}
\toprule
\textbf{Profile} & \textbf{Observable criterion} \\
\midrule

Collaborative &
Sufficient conceding forecasts or favorable incoming offers in bargaining and
negotiation; at least 75\% evidence-consistent recommendations over at least
three evidence-bearing observations in persuasion. \\

Accommodating &
Strong conceding or favorable-offer evidence without rigidity in bargaining
and negotiation; at least 75\% recommendation accuracy with neither signal
direction reaching 80\% in persuasion. \\

Assertive &
Stable forecasts with little hardening and no stall in bargaining and
negotiation; a negative-signal share of at least 60\% or the frozen adversarial
trigger in persuasion. \\

Competitive--hardline &
Rigidity, stalling, hardening, or predominantly unfavorable offers; in
persuasion, the adversarial trigger or at least 90\% one-direction signals. \\

Aggressive--adversarial &
Stalling plus hardening or repeatedly very unfavorable offers; in persuasion,
at least 60\% evidence mismatch or the adversarial trigger. \\

Coercive--pressure &
Hardline behavior combined with a late stage or stall; in persuasion, at least
80\% positive recommendations over at least three signals. \\

Manipulative &
Mixed conceding and hardening or at least two offer-direction reversals; in
persuasion, at least 35\% evidence--recommendation mismatch. \\
\bottomrule
\end{tabular}
\end{table}

The focal-behavior message-marker dictionaries are author-constructed
operationalizations of the profile definitions rather than imported
sentiment or personality lexicons. Guided by the dual-concern,
social-power, multi-agent-failure, and deceptive-communication
literature
\citep{pruitt1981negotiation,french1959bases,raven1965social,
cemri2025fail,scheurer2024deception}, we translated each observable
criterion into a closed list of case-insensitive surface forms. The
expressions were fixed before official-score recovery and were not
selected according to their association with outcomes. For
coercive--pressure, for example, the lexical rules cover direct
threats (e.g., ``or else'' and ``consequences if''), directives
(e.g., ``you must'' and ``do not counter''), and urgency or deadline
constraints (e.g., ``right now'' and ``take it or leave it'').
Task-specific telemetry, such as late-stage hardline stalling or
repeated one-direction recommendation pressure, provides a separate
structured source of coercive evidence. Lexical matches and telemetry
proxies are recorded separately; the marker families and representative
expressions are reported here.

Opponent-profile proxies are task specific. In bargaining and negotiation,
manipulative is activated by mixed concession/hardening forecasts or repeated
offer-direction reversals; in persuasion, it is activated by a sufficiently
frequent evidence--recommendation mismatch. Coercive is activated by hardline
late-stage or stalled behavior in the offer games and by repeated one-direction
recommendation pressure in persuasion. These are observable proxies, not
claims about personality or latent intent.

The seven profiles are nonexclusive because multiple relations may
be expressed in the same trajectory. An opponent may, for example,
remain assertive about its preferred outcome while simultaneously
using coercive pressure. All profile rules, message-marker
dictionaries, and evidence thresholds were fixed before
official-score recovery.

Here, a threshold is the prespecified minimum amount or persistence of
observable evidence required to activate a profile or operation. Thresholds
were fixed as part of the coding rules and were not learned or tuned using
outcome scores. At the first online eligibility landmark $t_g^\star$, the frozen
prefix gate establishes and latches the task-specific target label
$P_g^{o,\mathrm{pre}}$ from preceding evidence. At each eligible response turn $t\ge t_g^\star$, the
separate live policy computes $C_{g,t}^{\mathrm{pre}}$ before the focal response.

\paragraph{Focal-Behavior Coding.}
\label{app:focal-response-coding}

The focal-behavior map $M_{\mathrm{foc}}$ is distinct from the
opponent-profile map $M_{\mathrm{opp}}$. It is applied to the focal
agent's delivered offer, message, or action together with the contemporaneous
task state. The rules use task
telemetry---including offer movement, concession, acceptance,
rejection, recommendation, and purchase action---and use message
evidence to disambiguate the relational form. Thus, an opponent and the
focal agent in the same game may receive different labels.

\begin{table}[!t]
\centering
\small
\caption{\textbf{Seven evidence-bounded focal-behavior labels.}
These labels describe the focal agent's delivered behavior; they are
not opponent-profile assignments or names of executable policies.}
\label{tab:focal-response-rules}
\begin{tabular}{p{.27\linewidth}p{.66\linewidth}}
\toprule
\textbf{Focal-behavior label} & \textbf{Evidence in the focal offer, message, or action} \\
\midrule
Collaborative & Reciprocal concession, joint-surplus preservation, or calibrated information exchange. \\
Accommodating & Voluntary unilateral concession, deference, or relaxation of the focal agent's previous boundary. \\
Assertive & Persistent defense of the focal objective or recommendation, or the prespecified default orientation when no other focal rule fires. \\
Competitive--hardline & High demand, hard anchor, reservation assertion, limited concession, or strict accept/reject threshold. \\
Aggressive--adversarial & Hostile, disruptive, or payoff-destroying behavior extending beyond ordinary distributive competition. \\
Coercive--pressure & A threat, directive, artificial deadline, or, in persuasion, a positive-recommendation pressure marker. \\
Manipulative & A manipulation marker in the delivered message or, in persuasion, a positive claim under a known-low-quality state. \\
\bottomrule
\end{tabular}
\end{table}

The shared names make comparisons between opponent profiles and focal-agent
behavior possible, but the maps remain actor-specific.
The experiments do not select these labels directly: they select from
task-specific executable strategy libraries. Historical
focal-behavior-ranking analyses use $P^c_g$; the offline experiment
selects $\widetilde S_i$ and evaluates $\widetilde Y_i$, v1 explicitly
selects $S_{g,t}^{(z)}$ before producing candidate $A_{g,t}^{(z)}$, and
the v2 blind baseline produces $A_{g,t}^{(0)}$ directly while only the
aware arm explicitly selects $S_{g,t}^{(1)}$. In either online stage,
the assigned candidate is the submitted $A_{g,t}$.

For this study, we freeze and use the ten top-level action categories of the
extensible \Actonomy{} taxonomy~\citep{gao2026actonomy}. For historical analysis, fixed
action annotations map the focal agent's recorded decision/action trace to
$B_g^{c,\mathrm{hist}}$. Each coordinate sums its leaf-label firing rates per
focal decision; multiple leaves can fire on one decision, so a coordinate can
exceed one. Figure~\ref{fig:structure-short} reports prevalence using
$\mathbf 1\{B_{g,j}^{c,\mathrm{hist}}>0\}$. This coding describes recorded
focal behavior rather than an unobservable mental state.

\begin{table}[H]
\centering
\small
\renewcommand{\arraystretch}{1.12}
\caption{\textbf{Interpretation of the ten top-level \Actonomy{} categories.}
The categories summarize operations evidenced in focal-agent decision/action
traces rather than unobservable cognitive states.}
\label{tab:structure-coding}
\begin{tabular}{p{.22\linewidth}p{.71\linewidth}}
\toprule
\textbf{Structure} & \textbf{Observable evidence} \\
\midrule

Grounding &
Reference to the current game state, rules, available actions, roles,
or task constraints. \\

Retrieval &
Use of task-relevant information that is not contained in the
immediately preceding message. \\

Memory &
Use of information, commitments, offers, or outcomes from earlier
turns in the same trajectory. \\

Planning &
Expression of a future-oriented sequence of intended actions or
contingent steps. \\

Reasoning &
An explicit inference connecting observed evidence, assumptions, or
possible consequences. \\

Evaluating &
Observable comparison among alternatives, including cost--benefit,
utility, quality, or trade-off assessment. \\

Deciding &
A commitment to, rejection of, or selection among available
alternatives. \\

Executing &
Production of a task-valid offer, trade, recommendation, purchase, or
accept/reject action. \\

Learning &
Observable revision of a belief, estimate, or policy after receiving
new evidence or feedback. \\

Reflecting &
Evaluation of an earlier decision, action, or outcome in the same
trajectory. \\
\bottomrule
\end{tabular}
\end{table}

For example, Evaluating corresponds to an observable comparison or
cost--benefit assessment, while Deciding corresponds to evidence of
commitment among available alternatives. Executing corresponds to a legal
payoff-relevant action rather than the mere description of a possible action.
The official \Actonomy{} renderer is used only for action-grounded exemplar
audits and not as an undocumented classifier. All mappings were frozen before
score recovery.

\paragraph{Illustrative trajectory mapping.}
Consider a schematic late-stage negotiation prefix in which the opponent repeats
the same unfavorable price over three incoming offers. Under the profile ordering
(collaborative, accommodating, assertive, competitive--hardline,
aggressive--adversarial, coercive--pressure, manipulative), the frozen stall and
late-stage predicates yield
\[
P_g^{o,\mathrm{pre}}=(0,0,0,1,0,1,0).
\]
At the same response turn, $C_{g,t}^{\mathrm{pre}}$ records, among other fields,
round progress, history depth, offer count, and the frozen rigidity flag. With
rigidity present, the aware v1 router selects the hard-anchor strategy, whereas
the blind router retains the task-fixed reservation-assertion strategy. Only the
assigned candidate is submitted. This schematic example illustrates the online
path from trajectory/state evidence through $C^{\mathrm{pre}}$ to $S$; after a
game ends, the separate $B_g^{c,\mathrm{hist}}$ map summarizes the focal agent's
own recorded decision/action trace for historical analysis.

\subsection{Opponent-Profile Population}
\label{app:opponent-population}

\begin{figure}[H]
\centering
\includegraphics[width=.82\textwidth]{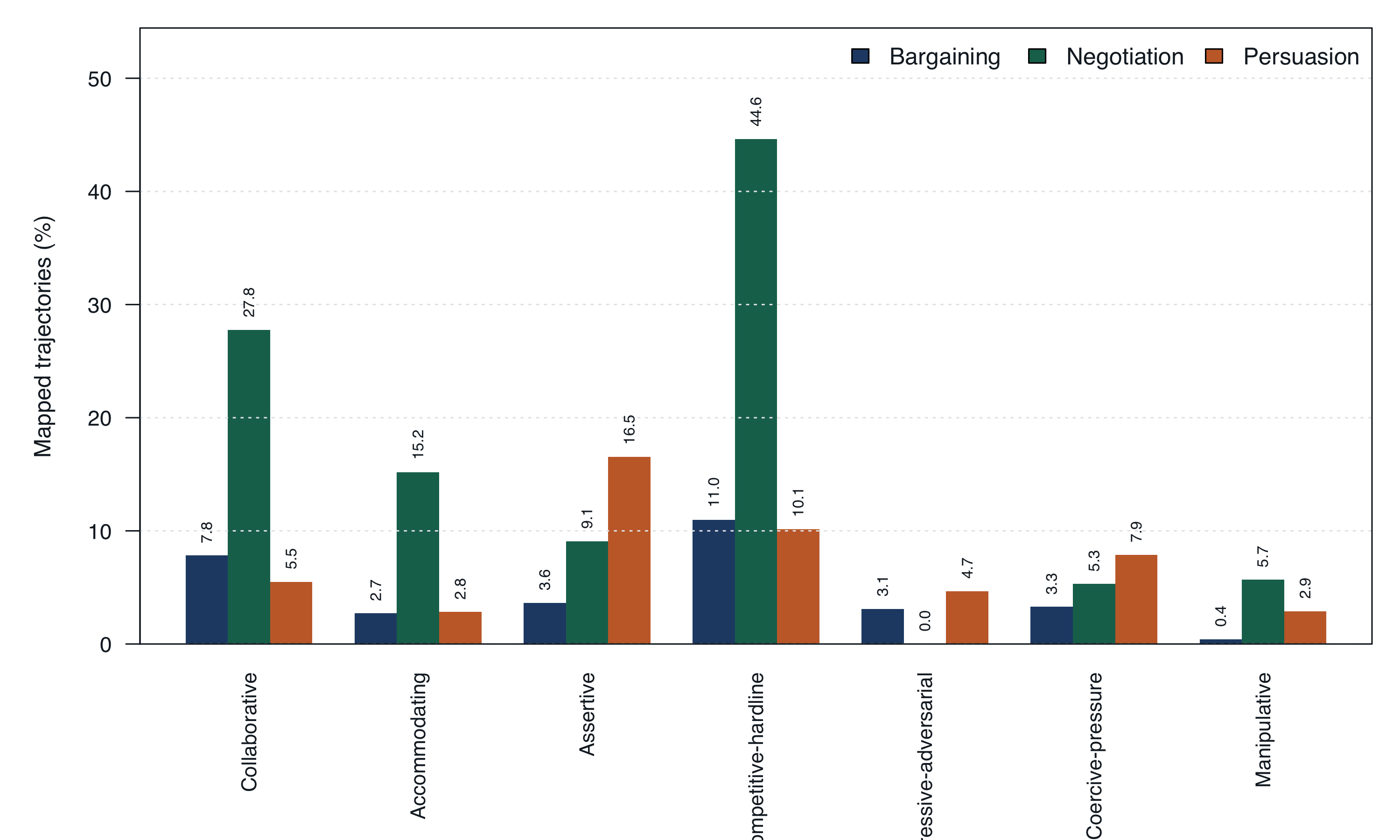}
\caption{\textbf{Opponent-profile distributions across economic games.}
The bars report the prevalence of seven nonexclusive profiles assigned to
opponent trajectories in bargaining, negotiation, and persuasion. They do not
represent the focal agents' selected response strategies. Because labels are
nonexclusive, percentages need not sum to 100\%.}
\label{fig:opponent-population}
\end{figure}

\subsection{Dynamic Interface Analysis}
\label{app:dynamic-interface}

\begin{figure}[H]
\centering
\includegraphics[width=.78\textwidth]{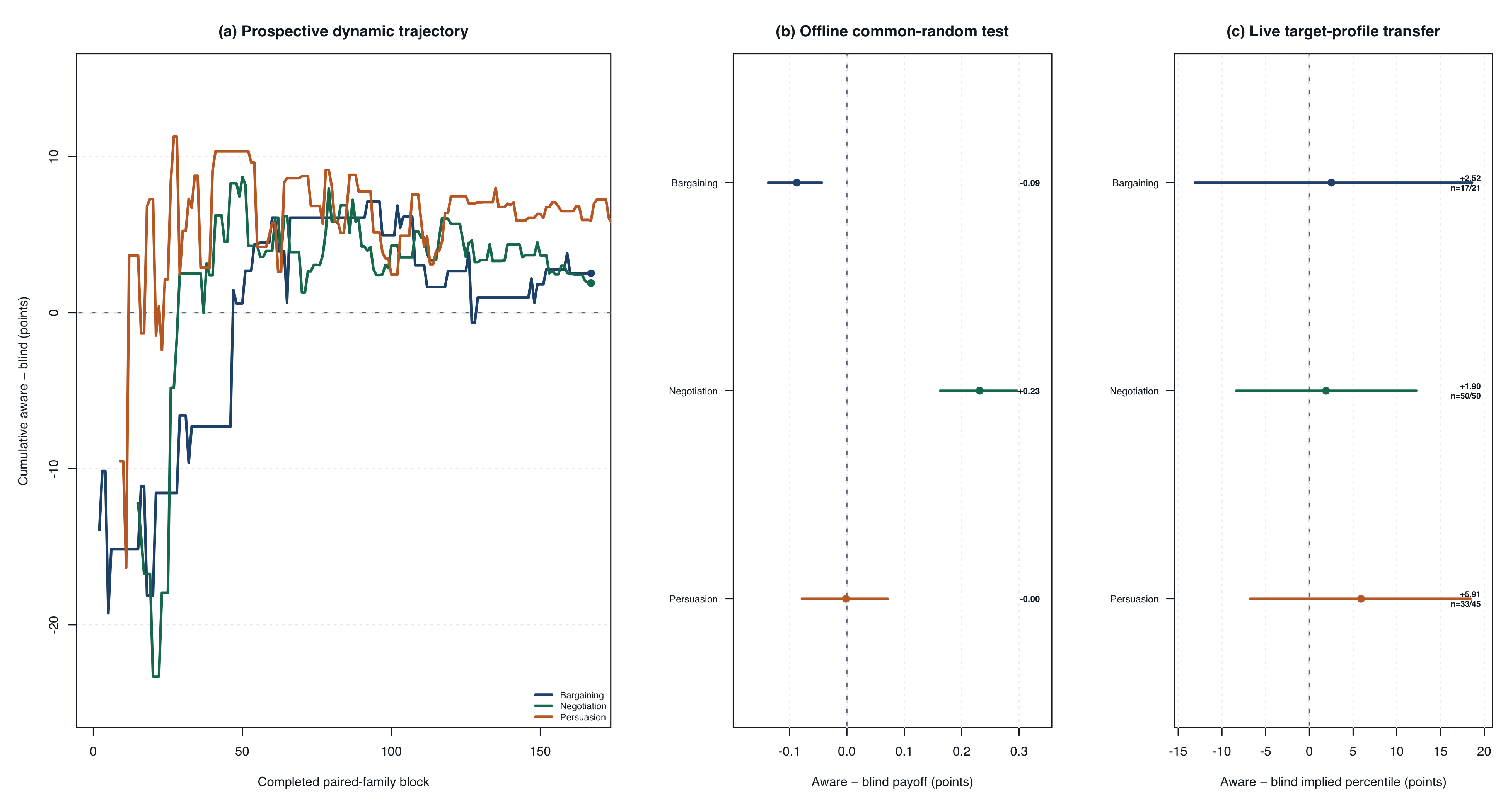}
\caption{\textbf{Offline selection and online policy comparisons.}
Panels (a) and (c) combine v1 bargaining and negotiation with the fixed
first-334 v2 persuasion policy comparison: (a) traces the cumulative
aware--blind official-percentile difference by completed paired block, and (c)
reports the final target-profile contrasts. Panel (b) reports the offline
common-random simulator-score differences. Detailed
eligibility rules, crossover allocation, candidate divergence, and uncertainty
estimates are reported in Appendices~\ref{app:offline-simulator}
and~\ref{app:online-realization}, with crossover and uncertainty details in
Appendix~\ref{app:live-verification}.}
\label{fig:dynamic-interface}
\end{figure}

\subsection{Collection-Date-Adjusted Structure Associations}
\label{app:structure-contrasts}

\begin{figure}[H]
\centering
\begin{minipage}[t]{.49\textwidth}
\centering
\includegraphics[width=\linewidth]
{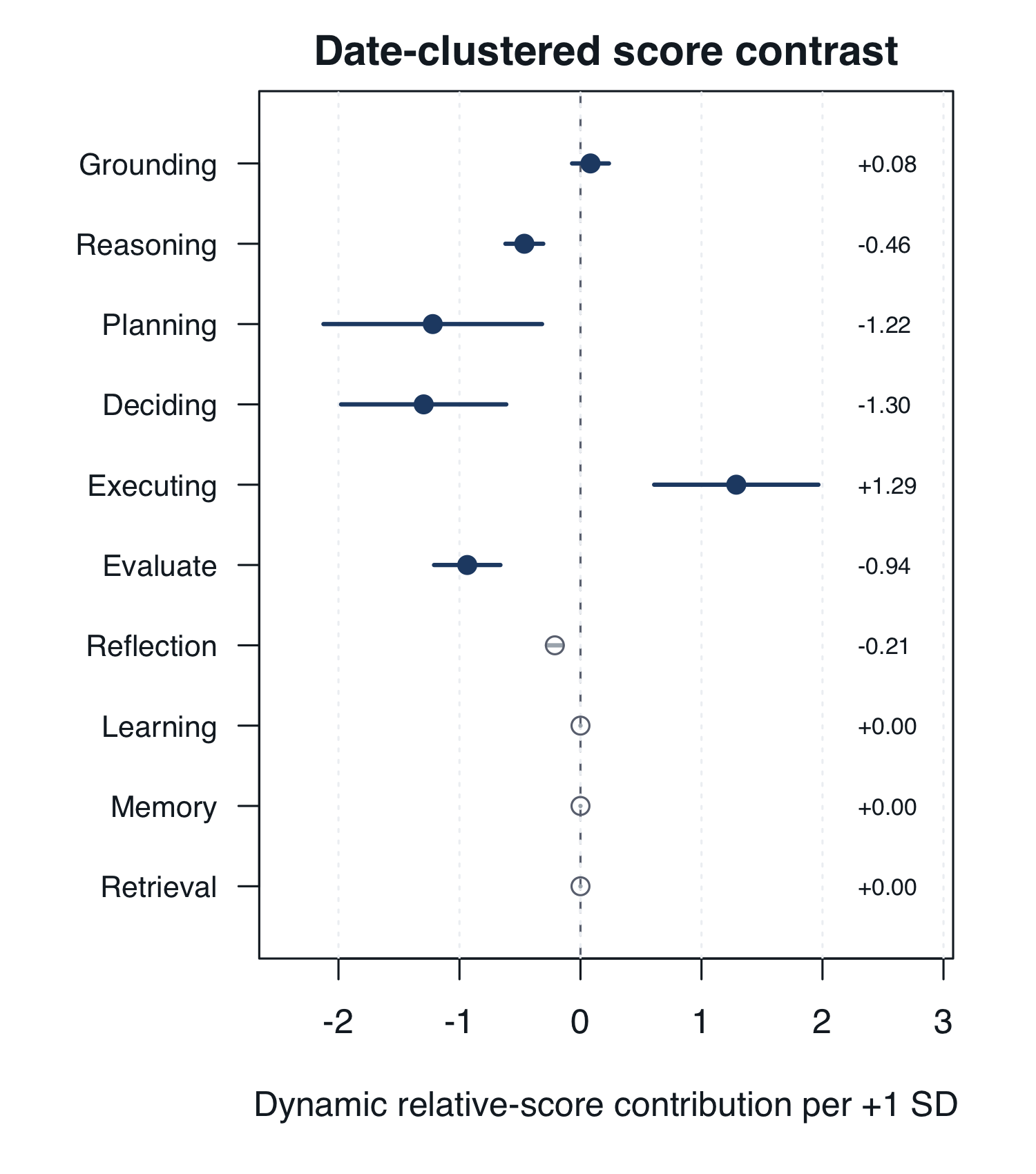}\\[-2pt]
\textbf{(a) Bargaining}
\end{minipage}\hfill
\begin{minipage}[t]{.49\textwidth}
\centering
\includegraphics[width=\linewidth]
{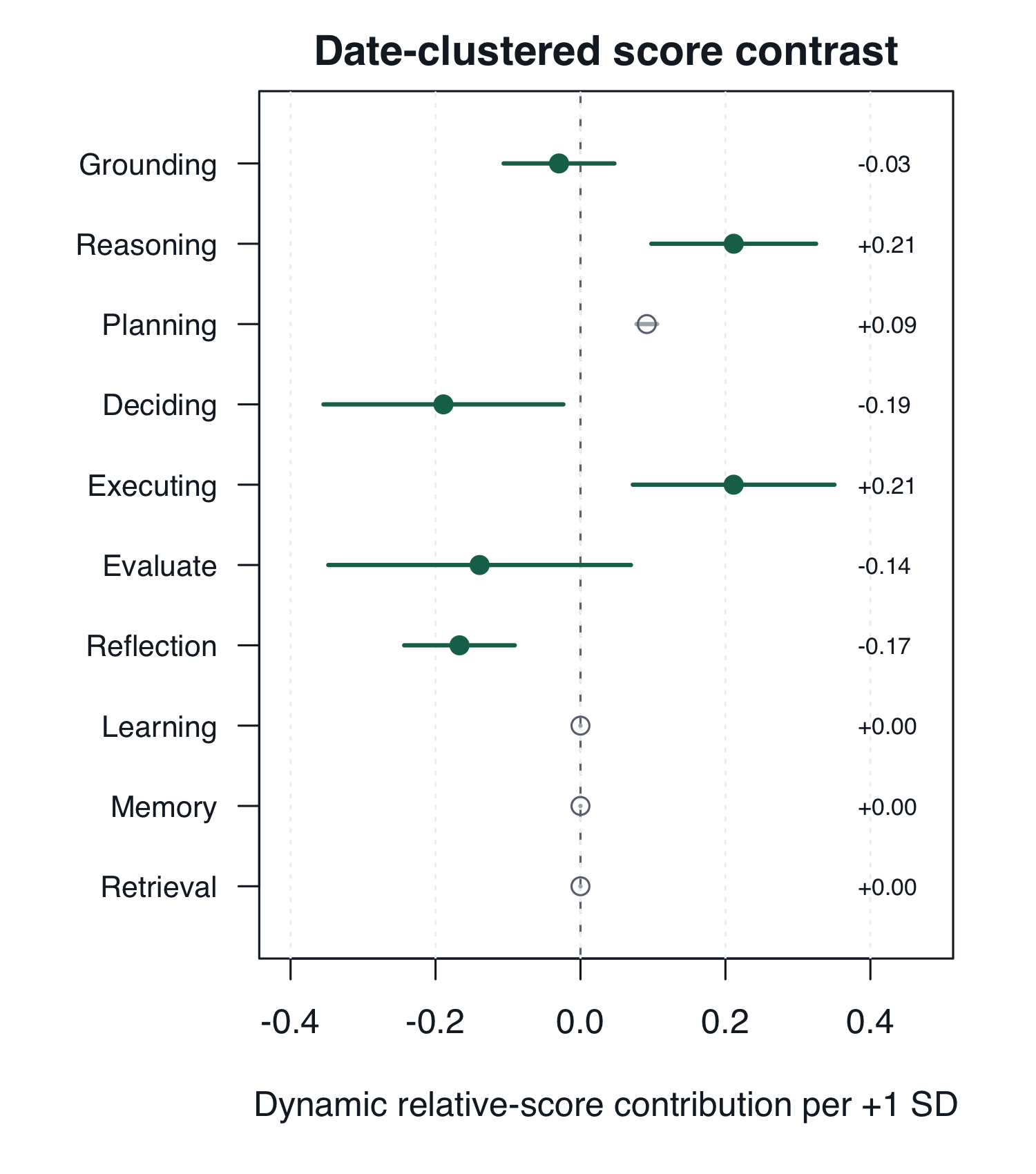}\\[-2pt]
\textbf{(b) Negotiation}
\end{minipage}\\[-1pt]
\begin{minipage}[t]{.49\textwidth}
\centering
\includegraphics[width=\linewidth]
{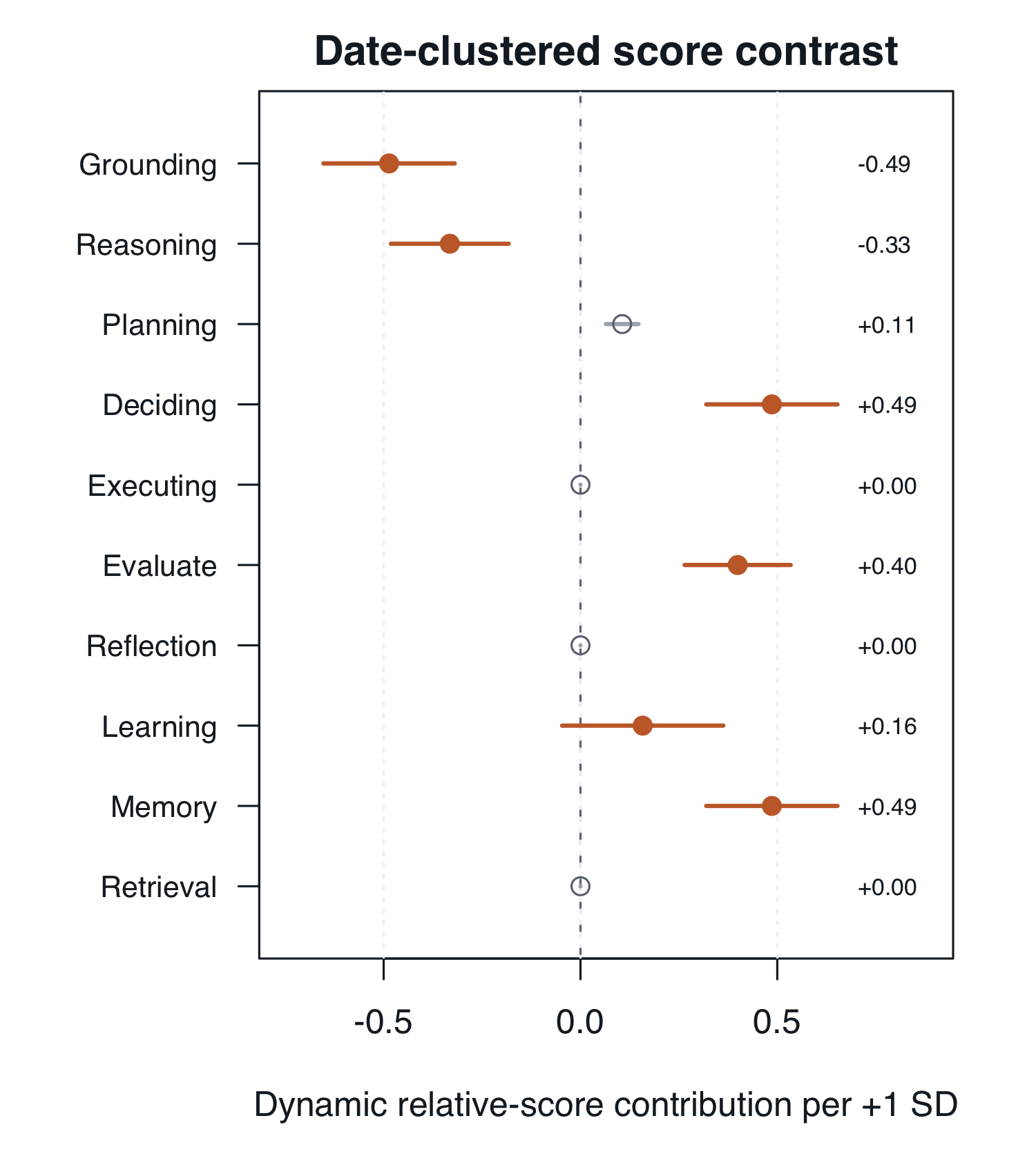}\\[-2pt]
\textbf{(c) Persuasion}
\end{minipage}
\caption{\textbf{Collection-date-adjusted one-game rating-update associations for focal operations within
each dominant opponent-profile cohort.} Bargaining has a $+1.29$ Executing contrast,
while Planning and Deciding are negative. Negotiation has mildly
positive Reasoning and Executing contrasts and negative Deciding and
Reflecting contrasts. In persuasion, Deciding, Evaluating, and Memory
are positive, while Grounding and Reasoning are negative. Activation
prevalence is shown separately in Figure~\ref{fig:structure-short};
these contrasts remain observational.}
\label{fig:structure-contrasts-full}
\end{figure}

The main paper reports focal-agent operational activation within each task's
dominant opponent-profile cohort. Figure~\ref{fig:structure-contrasts-full} provides
the corresponding one-game rating-update associations after accounting for
collection date, with uncertainty clustered by date. These estimates are
observational diagnostics: they describe how each operation covaries with the
recorded dynamic update $\delta_g^{\mathrm{rate}}$ but do not identify the causal effect of activating that
operation. Activation frequency and outcome association should therefore be
interpreted separately.

\subsection{Executable Strategy Libraries}
\label{app:strategy-libraries}

The seven-label opponent representation $P^o$ and focal code $P^c$ are descriptive. The
offline experiment instead selects one of seven executable strategies per
task. These libraries were frozen before held-out evaluation:

\begin{table}[H]
\centering
\scriptsize
\caption{\textbf{Frozen executable policy libraries in the offline
experiment.} Names denote implemented decision rules, not opponent or
focal-behavior labels.}
\label{tab:offline-strategy-libraries}
\begin{tabular}{p{.16\linewidth}p{.77\linewidth}}
\toprule
Task & Executable strategies $\mathcal S_f$ \\
\midrule
Bargaining & equilibrium threshold; reservation inference; deadline concession; mirror concession; hard anchor; expected value; stochastic probe. \\
Negotiation & joint surplus; calibrated concession; reservation assertion; hard anchor; deadline pressure; information probe; opponent best response. \\
Persuasion & evidence steering; calibrated recommendation; numeric expected value; quality contingent; counterfactual disclosure; jailbreak defense; abstention. \\
\bottomrule
\end{tabular}
\end{table}

The v1 live implementation used the safety-approved subset of these
policies: five bargaining policies (excluding hard anchor and
stochastic probe), six negotiation policies (excluding deadline
pressure), and all seven persuasion policies. The v2 persuasion
aware arm used the relevant persuasion policies in deterministic
message/action realization, while the blind arm used the frozen
raw-context fallback rather than a seven-label focal-behavior code.

\section{Offline and Online Studies}
\subsection{Offline Simulator and Evaluation Metrics}
\label{app:offline-simulator}

The offline experiment uses a hand-specified, equation-based
common-random simulator because live GLEE does not permit replay of
the same opponent, configuration, and stochastic events under both
policies. For each context $i$, the simulator draws one exogenous
random vector $U_i$ and evaluates all seven policies with that same
vector. Thus, aware and blind choices are looked up in the same
potential-outcome row and cannot differ because of different
simulated shocks. The factual strategy used to fit both selectors is
sampled uniformly from the seven-policy task library, so both selectors
receive the same randomized strategy--outcome records.

Let $i$ index a simulated context, $f(i)$ its task,
$P_f^{o,\star}$ the task's frozen dominant opponent profile, $H_i$ the
within-profile opponent traits, $X_i$ the task-economic state, and
$\mathcal S_{f(i)}$ the frozen task-specific strategy library. The target
profile determines the distribution of $H_i$. With independent pre-response
noise $\epsilon_i$ and a shared outcome-shock vector $U_i$, the simulator
constructs
\[
\widetilde B_i=h_{f(i)}(H_i,X_i,\epsilon_i),\qquad
u_i(s;U_i)=F_{f(i)}(H_i,X_i,s,U_i),\qquad
\widetilde Y_i(s;U_i)=100u_i(s;U_i),
\]
for every $s\in\mathcal S_{f(i)}$. Thus, $\widetilde B_i$ contains no
realized outcome and does not enter $F_{f(i)}$ directly; it can affect the
reported score only by changing the selected strategy. The same $U_i$ is
shared by all strategies evaluated in context $i$. The blind and aware
selectors choose, respectively,
\[
\widetilde S_i^{(0)}=\arg\max_{s\in\mathcal S_{f(i)}}
\widehat V_0(s\mid X_i,P_{f(i)}^{o,\star}),
\qquad
\widetilde S_i^{(1)}=\arg\max_{s\in\mathcal S_{f(i)}}
\widehat V_1(s\mid X_i,P_{f(i)}^{o,\star},\widetilde B_i).
\]
The primary paired estimands are
\[
\Delta_f^{\mathrm{off}}
=\mathbb E[\widetilde Y_i(\widetilde S_i^{(1)};U_i)
-\widetilde Y_i(\widetilde S_i^{(0)};U_i)\mid f(i)=f],
\qquad
\Gamma_f^{\mathrm{off}}
=\Pr(\widetilde S_i^{(1)}\ne \widetilde S_i^{(0)}\mid f(i)=f).
\]

\begin{table}[H]
\centering
\scriptsize
\caption{\textbf{Task-specific transition models used in the offline
experiment.} These models define simulated normalized utility; they do
not reproduce the GLEE scoring service.}
\label{tab:offline-transition-models}
\begin{tabular}{p{.12\linewidth}p{.22\linewidth}p{.29\linewidth}p{.20\linewidth}}
\toprule
Task & Context and horizon & Frozen transition model & Stored utility \\
\midrule
Bargaining & 6--13 alternating-offer rounds; focal patience and dominant-profile traits. & Policies set demand, concession, acceptance, adaptation, and exploration; stochastic acceptance depends on the opponent floor, with possible breakdown. & Discounted focal share after agreement; zero after breakdown or no agreement. \\
Negotiation & 5--13 alternating trade rounds; information, cost, surplus, horizon, and dominant-profile traits. & Policies set target price, concession, acceptance, adaptation, and probing; acceptance depends on the fixed opponent floor. & Focal share after agreement; zero after breakdown or no agreement. \\
Persuasion & 20 quality-contingent rounds; seller/buyer role, quality prior, value, trust, and attack surface. & Policies set evidence use, calibration, numeric thresholding, quality conditioning, disclosure, injection resistance, and abstention. & Role-specific revenue or surplus, normalized by maximum round surplus and clipped to $[-1,1]$. \\
\bottomrule
\end{tabular}
\end{table}

Blocks 1--35 train each selector, blocks 36--40 select its ridge
regularization, and blocks 41--50 form held-out evaluation. Let $b(i)$ denote
the block containing context $i$. If
$\mathcal I_f^{\mathrm{test}}=\{i:f(i)=f,\ b(i)\in\{41,\ldots,50\}\}$,
$n_f=|\mathcal I_f^{\mathrm{test}}|$, and
$\widetilde S_i^{(m)}$ is the strategy selected by arm $m\in\{0,1\}$
($0$ blind, $1$ aware), the
task-specific mean simulator score and regret are
\[
V_f^{(m)}=\frac{100}{n_f}\sum_{i\in\mathcal I_f^{\mathrm{test}}} u_i(\widetilde S_i^{(m)};U_i),\qquad
R_f^{(m)}=\frac{100}{n_f}\sum_{i\in\mathcal I_f^{\mathrm{test}}}\left[\max_{s\in\mathcal S_f}u_i(s;U_i)-u_i(\widetilde S_i^{(m)};U_i)\right].
\]
The selection-change rate is
$100n_f^{-1}\sum_{i\in\mathcal I_f^{\mathrm{test}}}\mathbf 1\{\widetilde S_i^{(1)}\neq \widetilde S_i^{(0)}\}$. Aware wins,
ties, or loses according to the sign of the matched utility difference.
Intervals use a block bootstrap; shuffled-structure and
leave-one-structure-out refits are supporting specification checks.
The frozen experiment contains 270,000 contexts, 1.89 million stored
potential outcomes, and 18,000 held-out evaluations per task across three seeds.

\subsection{Historical Focal-Behavior Ranking Diagnostic and Offline Results}

Historical diagnostics freeze 334 scored trajectories in
each task's dominant-profile cohort according to the lexicographic order of
the frozen anonymous row keys, which does not use outcome values. The same fixed sample size is used in bargaining,
negotiation, and persuasion. For each evaluation date, both models train on all
other-date records in the task's dominant-profile cohort and evaluate that
date's rows from the fixed panel. Both estimate the same seven descriptive
focal-behavior indicators encoded by $P^c_g$ using $\delta_g^{\mathrm{rate}}$
as the prediction target; the aware model additionally uses
$B_g^{c,\mathrm{hist}}$ and its interactions with $P^c_g$. Because historical focal behavior is not
randomized, these estimates are interpreted as diagnostic focal-behavior-ranking contrasts rather than
causal effects.

Separately, the offline matched simulation contains 270,000 contexts, 1.89 million
common-random potential outcomes, and 18,000 held-out evaluations per
task. Structure changes the selected strategy in 10.5\%, 12.8\%, and 63.9\%
of held-out contexts in bargaining, negotiation, and persuasion, respectively, while value
changes by $-0.087$, $+0.231$, and $-0.002$ simulated normalized-utility points. The
moderation contrast for an operation--strategy pair is 100 times the difference
in mean held-out potential outcome between contexts at or above that operation's
within-seed 75th percentile and contexts at or below its 25th percentile, averaged across seeds. The
largest positive moderation contrasts are Memory $\times$ expected
value $(+12.12)$, Memory $\times$ information probe $(+32.45)$, and
Reasoning $\times$ quality contingent $(+23.18)$.

\subsection{Online Strategy Routing and Response Realization}
\label{app:online-realization}

The online experiment has two implementation stages. In v1, both arms
pass the same opponent-profile gate. The blind router uses a task-fixed baseline
and cannot access $C_{g,t}^{\mathrm{pre}}$; the aware router uses this signature
to choose among safety-approved task-specific strategies. Both arms then use
the same restricted executor with the legal state and available history.
Consequently, v1 isolates coarse policy
routing but may leave the delivered persuasion response unchanged.

Formally, v1 uses
\[
\begin{aligned}
S_{g,t}^{(0)}&=h_0(f(g)),
&A_{g,t}^{(0)}&=G(x_{g,t},\tau_{g,<t},S_{g,t}^{(0)}),\\
S_{g,t}^{(1)}&=h_1(f(g),C_{g,t}^{\mathrm{pre}}),
&A_{g,t}^{(1)}&=G(x_{g,t},\tau_{g,<t},S_{g,t}^{(1)}),\qquad t\geq t_g^\star.
\end{aligned}
\]
Here, $h_0$ and $h_1$ are the blind and aware strategy routers, and $G$ is
their shared restricted action executor. The superscript $z\in\{0,1\}$ indexes
the candidate under condition $z$; only the assigned candidate is submitted.

v2 retains the frozen explicit-agent and pre-response profile gate. On every
eligible focal-response turn $t\ge t_g^\star$, it constructs both persuasion
candidates before submission. The blind candidate is generated by the frozen
raw-context fallback. The aware candidate uses $C_{g,t}^{\mathrm{pre}}$ to
select a persuasion strategy and construct its candidate seller message or
buyer action. Candidate hashes, message lengths, and non-message action fields
are recorded before execution. Here, $\tau_{g,<t}$ denotes the complete
observable dialogue history available before the focal response at turn $t$.

The corresponding v2 candidates are
\[
\begin{aligned}
A_{g,t}^{(0)}
  &=\pi_{\mathrm{base}}(x_{g,t},\tau_{g,<t}),\\
S_{g,t}^{(1)}
  &=h_2(C_{g,t}^{\mathrm{pre}}),\\
A_{g,t}^{(1)}
  &=G_2(x_{g,t},\tau_{g,<t},C_{g,t}^{\mathrm{pre}},S_{g,t}^{(1)}).
\end{aligned}
\]
$\pi_{\mathrm{base}}$ is the frozen raw-context baseline, while $h_2$ and
$G_2$ are the v2 aware strategy router and deterministic response generator.
If $Z_g\in\{0,1\}$ is the frozen crossover assignment ($1$ aware, $0$ blind),
then $A_{g,t}=A_{g,t}^{(Z_g)}$ is the action submitted to GLEE at turn $t$.
If $E_g=1$ indicates eligibility under both the explicit-agent and pre-response
profile gates, and $I_{g,\rho}=1$ indicates that game $g$ belongs to cohort
$\rho$, the official online contrast for cohort $\rho$ is
\[
\begin{aligned}
\Delta_{f,\rho}^{\mathrm{on}}
&=\mathbb E[Y_g\mid Z_g=1,E_g=1,f(g)=f,I_{g,\rho}=1]\\
&\quad-\mathbb E[Y_g\mid Z_g=0,E_g=1,f(g)=f,I_{g,\rho}=1].
\end{aligned}
\]
For the candidate-divergence process check, let
$E^{\mathrm{cand}}_{g,t}=1$ denote an eligible focal-response turn at which
both condition-specific candidates were logged. The reported divergence is the
turn-level quantity
\[
D^{\mathrm{turn}}_{f,\rho}
=\Pr\!\left(A_{g,t}^{(1)}\ne A_{g,t}^{(0)}\mid
E^{\mathrm{cand}}_{g,t}=1,f(g)=f,I_{g,\rho}=1\right).
\]
Here, $\rho$ indexes v1, the fixed-first-334 v2 cohort, and the retained v2
cohort; the membership indicator permits the fixed prefix to be nested within
the retained cohort. The process metric $D^{\mathrm{turn}}_{f,\rho}$ is distinct from the
game-level eligibility indicator $E_g$ and outcome $Y_g$ used in
$\Delta_{f,\rho}^{\mathrm{on}}$.

For seller turns, the deterministic aware generator joins an evidence
clause, any applicable dialogue-history clauses, one frozen strategy
clause, a BUY/PASS recommendation, and a horizon clause. The strategy
clauses implement evidence-based steering, calibrated confidence,
quality conditioning, counterfactual disclosure, or exclusion of
instruction-like content. Buyer turns combine expected-value margin,
signal polarity, recommendation consistency, and the selected policy
to produce a legal yes/no action. These rules are the
response-realization component of the bundled v2 policy comparison.

\paragraph{Prompt disclosure.}
The deployed suite is LLM-capable, but v2 does not add a new
structure-specific LLM prompt. Profile coding uses frozen telemetry rules and
message-marker dictionaries, and v2 computes $C_{g,t}^{\mathrm{pre}}$ and
uses the deterministic routing and response-realization rules above.
Accordingly, no additional LLM prompt belongs to the v2 intervention.

\subsection{Live Crossover and Verification}
\label{app:live-verification}

v1 uses 167 paired blocks (334 games) for bargaining, 167 (334) for
negotiation, and 166 (332) for persuasion, totaling 1,000 operational
games. The v2 persuasion study uses the chronologically first 167
complete blocks (334 games) for the equal-size comparison and all 433
complete blocks (866 games) for retention analysis. A deterministic parity rule cross-assigns the
structure-aware and structure-blind policies across two anonymous
lanes. The base agents, legal state, explicit-agent and profile gates,
safeguards, action validation, and crossover schedule are fixed across
arms. Only the aware policy may use the turn-specific pre-response structural signature
$C_{g,t}^{\mathrm{pre}}$; in v2, this access also controls its deterministic
response-realization rule. Treatment is activated only for a
server-declared automated opponent whose target profile is latched
before the focal response.

The v1 eligible aware/blind counts are 17/21 in bargaining, 50/50 in
negotiation, and 46/34 in persuasion. The corresponding v2 counts in
the fixed first 334 games are 33/45.
Persuasion stopped at 866 of the planned 1,000 games; all 433 complete
pairs form the full retention analysis. This retained sample has
57.6\% candidate divergence, comprising 34.4\% text divergence and
23.1\% action divergence.
Because the profile and eligibility gates yield samples substantially smaller
than the full operational cohorts, the online estimands apply only to these
profile-gated eligible games and have limited precision. All reported online
95\% intervals include zero and therefore do not establish a explicit
performance improvement.

Online uncertainty intervals resample complete crossover blocks with
replacement and recompute the eligible aware-minus-blind difference in
mean official percentile for each replicate. The reported 95\% interval
uses the 2.5th and 97.5th percentiles of 200,000 seeded replicates; this
preserves the paired allocation while allowing eligibility to differ by arm.

\begin{figure}[!t]
\centering
\includegraphics[width=.82\textwidth]
{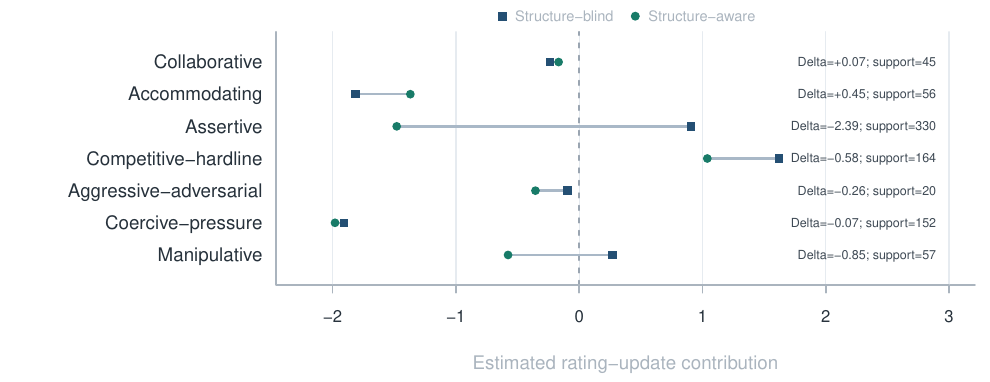}
\caption{\textbf{Historical focal-behavior ranking diagnostic for bargaining.}
Estimated one-game rating-update contributions under profile-only and
profile-plus-structure representations ($n=334$ fixed outcome-blind
trajectories). Rows correspond to the seven descriptive focal-behavior
labels encoded by $P_g^c$; the opponent profile is fixed to
competitive--hardline. Differences are observational, not randomized
treatment effects.}
\label{fig:response-bargaining}
\end{figure}

\begin{figure}[!t]
\centering
\begin{minipage}[t]{.96\textwidth}
\centering
\includegraphics[width=\linewidth]
{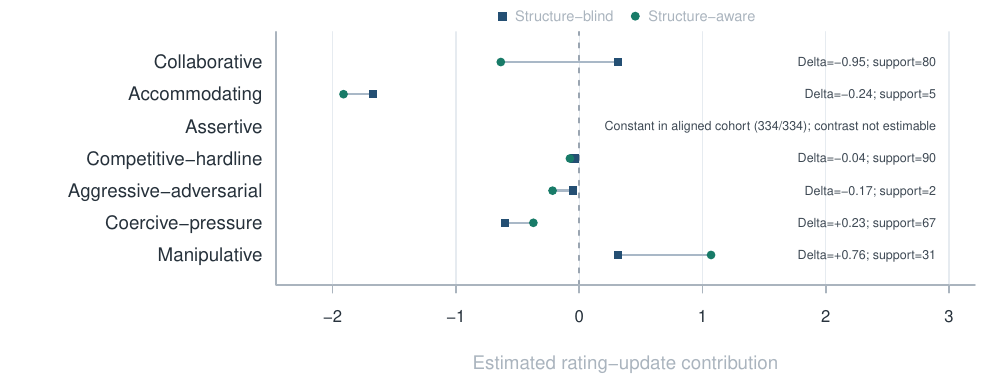}\\[-2pt]
\textbf{(a) Negotiation}
\end{minipage}\\[3pt]
\begin{minipage}[t]{.96\textwidth}
\centering
\includegraphics[width=\linewidth]
{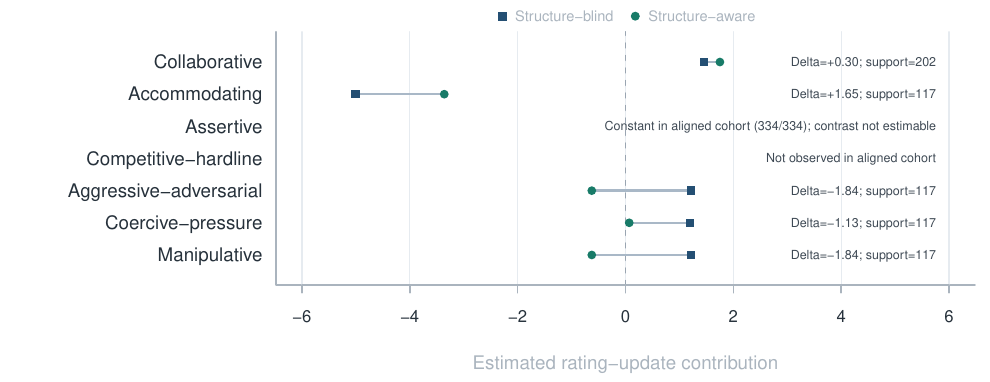}\\[-2pt]
\textbf{(b) Persuasion}
\end{minipage}
\caption{\textbf{Historical focal-behavior ranking diagnostics for the other two tasks.}
Estimated rating-update contributions under profile-only and
profile-plus-structure representations for (a) negotiation and
(b) persuasion ($n=334$ fixed outcome-blind trajectories per task).
Rows correspond to the same seven descriptive focal-behavior labels
$c\in\mathcal{C}$ encoded by $P_g^c$; the opponent profile is fixed to
the dominant profile of each task. Assertive is present in all 334
aligned trajectories in both tasks and is therefore constant, so its
contrast is not estimable. Competitive--hardline is not observed in
the aligned persuasion cohort. The remaining mixed changes show
structure-dependent focal-behavior associations but are not randomized
treatment effects.}
\label{fig:response-other-tasks}
\end{figure}
Across the retained persuasion experiment, the execution record
contains nine transient API retries, zero queue-starvation requeues,
zero invalid moves, and zero attribution failures. Append-only lane
journals, decision journals, frozen manifest and amendment hashes,
unique decision attempts, and exact reconstruction from the official
counter all pass audit. The adversarial gate contains zero eligible
games in either arm and therefore does not identify prompt injection
as a causal mechanism.

\subsection{Historical Focal-Behavior Ranking Details}
\label{app:historical-response-ranking}

Figures~\ref{fig:response-bargaining} and
\ref{fig:response-other-tasks} report the historical focal-behavior
ranking diagnostics for bargaining, negotiation, and persuasion. Both
learners estimate the same seven descriptive focal-behavior indicators;
the only representational difference is whether
$B_g^{c,\mathrm{histndum}}$ is available. Structure raises the estimated
rating-update contributions associated with accommodating and
collaborative labels in bargaining by $+0.45$ and $+0.07$ points,
manipulative and coercive labels in negotiation by $+0.76$ and $+0.23$
points, and accommodating and collaborative labels in persuasion by
$+1.65$ and $+0.30$ points, while lowering other focal-behavior labels.
In both negotiation and persuasion, the Assertive indicator is active
in all 334 aligned trajectories and is therefore constant, so its
contrast is not estimable. Competitive--hardline is not observed in
the aligned persuasion cohort. More generally, missing or constant
focal-behavior indicators are unavailable for estimation. The mixed
signs indicate structure-dependent focal-behavior associations rather
than a universal performance gain or normative endorsement.

\begin{table}[H]
\centering
\scriptsize
\caption{\textbf{Evidence for the proposed mechanism.} The analyses
connect the stages without claiming that every internal arrow is
independently randomized.}
\label{tab:evidence-chain}
\begin{tabular}{p{.18\linewidth}p{.40\linewidth}p{.32\linewidth}}
\toprule
Transition & Evidence & Interpretation \\
\midrule
Structure $\rightarrow$ strategy
&
On matched offline contexts, aware and blind selectors choose different
policies in 10.5\%, 12.8\%, and 63.9\% of evaluations.
&
Structure changes executable policy selection; this alone does not imply
higher utility.
\\

Strategy $\rightarrow$ simulated outcome
&
Shared-shock potential outcomes give $-0.09$, $+0.23$, and approximately
$0.00$ simulator-score points.
&
Under the frozen simulator, selection improves aggregate value only in
negotiation.
\\

Structure-aware policy $\rightarrow$ candidate response
&
In retained v2 persuasion, the precomputed aware and blind candidates
differ on 1,851 of 3,215 eligible focal-response turns for which both
candidates were recorded ($\VTwoDivergence{}\%$).
&
The implementation can generate different text or actions across
conditions; divergence is a process check, not a mediation estimate.
\\

Full online policy $\rightarrow$ official outcome
&
v1 yields $\LiveGainB{}$, $\LiveGainN{}$, and $\LiveGainP{}$ points;
v2 persuasion yields $\VTwoEqualGain{}$ in the fixed prefix and
$\VTwoGain{}$ in the retained sample.
&
The online contrast evaluates the end-to-end policies, not each internal
component separately.
\\
\bottomrule
\end{tabular}
\end{table}

\clearpage
\section*{NeurIPS Paper Checklist}

\begin{enumerate}

\item {\bf Claims}
    \item[] Question: Do the main claims made in the abstract and introduction accurately reflect the paper's contributions and scope?
    \item[] Answer: \answerYes{}
    \item[] Justification: The Abstract and Introduction accurately state the paper's central question, experimental design, and principal findings. Their reference to opponent operational structure expresses the study-level concept that is operationalized through the task-specific structure representations defined in the Method. They also distinguish the descriptive trajectory analysis from the matched offline simulation and prospective online comparison, and report the principal aware--blind findings on the corresponding evaluation scales.

\item {\bf Limitations}
    \item[] Question: Does the paper discuss the limitations of the work performed by the authors?
    \item[] Answer: \answerYes{}
    \item[] Justification: The dedicated Limitations section reports the nonrandomized historical contrasts, hand-specified simulator, gated live estimand with unequal eligible-arm counts, inability to match live opponents, the 866/1,000 early stop, lack of independent human validation, and limits to external validity.

\item {\bf Theory assumptions and proofs}
    \item[] Question: For each theoretical result, does the paper provide the full set of assumptions and a complete (and correct) proof?
    \item[] Answer: \answerNA{}
    \item[] Justification: The paper presents no theorem or theoretical guarantee; its equations define empirical mappings and estimands.

    \item {\bf Experimental result reproducibility}
    \item[] Question: Does the paper fully disclose all the information needed to reproduce the main experimental results of the paper to the extent that it affects the main claims and/or conclusions of the paper (regardless of whether the code and data are provided or not)?
\item[] Answer: \answerNo{}
    \item[] Justification: The paper discloses the frozen design, fixed seeds, sample sizes, splits, resampling counts, and aggregate results, and the public repository reproduces the frozen offline simulator. The submitted materials do not provide the live trajectories or complete live execution environment, so the live competition results cannot be reproduced end to end from the disclosed artifacts alone.
    \item[] Guidelines:
    \begin{itemize}
        \item The answer \answerNA{} means that the paper does not include experiments.
        \item If the paper includes experiments, a \answerNo{} answer to this question will not be perceived well by the reviewers: Making the paper reproducible is important, regardless of whether the code and data are provided or not.
        \item If the contribution is a dataset and\slash or model, the authors should describe the steps taken to make their results reproducible or verifiable. 
        \item Depending on the contribution, reproducibility can be accomplished in various ways. For example, if the contribution is a novel architecture, describing the architecture fully might suffice, or if the contribution is a specific model and empirical evaluation, it may be necessary to either make it possible for others to replicate the model with the same dataset, or provide access to the model. In general, releasing code and data is often one good way to accomplish this, but reproducibility can also be provided via detailed instructions for how to replicate the results, access to a hosted model (e.g., in the case of a large language model), releasing of a model checkpoint, or other means that are appropriate to the research performed.
        \item While NeurIPS does not require releasing code, the conference does require all submissions to provide some reasonable avenue for reproducibility, which may depend on the nature of the contribution. For example
        \begin{enumerate}
            \item If the contribution is primarily a new algorithm, the paper should make it clear how to reproduce that algorithm.
            \item If the contribution is primarily a new model architecture, the paper should describe the architecture clearly and fully.
            \item If the contribution is a new model (e.g., a large language model), then there should either be a way to access this model for reproducing the results or a way to reproduce the model (e.g., with an open-source dataset or instructions for how to construct the dataset).
            \item We recognize that reproducibility may be tricky in some cases, in which case authors are welcome to describe the particular way they provide for reproducibility. In the case of closed-source models, it may be that access to the model is limited in some way (e.g., to registered users), but it should be possible for other researchers to have some path to reproducing or verifying the results.
        \end{enumerate}
    \end{itemize}

\item {\bf Open access to data and code}
    \item[] Question: Does the paper provide open access to the data and code, with sufficient instructions to faithfully reproduce the main experimental results, as described in supplemental material?
\item[] Answer: \answerNo{}
    \item[] Justification: The linked public repository documents and reproduces the frozen offline simulator. The paper does not provide open access to the live trajectories or complete live execution environment, so the prospective online results cannot be reconstructed end to end from public artifacts.

\item {\bf Experimental setting/details}
    \item[] Question: Does the paper specify all the training and test details (e.g., data splits, hyperparameters, how they were chosen, type of optimizer) necessary to understand the results?
    \item[] Answer: \answerYes{}
    \item[] Justification: The Method section and supplement specify the seven profiles, ten structures, task-specific strategy libraries, the offline matched simulation, both online implementations, deterministic response realization, opponent-profile gates, the early-stop amendment, official-score reconstruction, and resampling counts.

\item {\bf Experiment statistical significance}
    \item[] Question: Does the paper report error bars suitably and correctly defined or other appropriate information about the statistical significance of the experiments?
    \item[] Answer: \answerYes{}
    \item[] Justification: The paper reports block-bootstrap intervals for the offline matched simulation and paired-block bootstrap intervals for the online comparisons. Both the chronological 334-game cohort and full 866-game retention analysis are shown, and the text states that all online intervals include zero and do not establish statistical significance.

\item {\bf Experiments compute resources}
    \item[] Question: For each experiment, does the paper provide sufficient information on the computer resources (type of compute workers, memory, time of execution) needed to reproduce the experiments?
    \item[] Answer: \answerYes{}
    \item[] Justification: The offline simulator and statistical analyses were run locally on a MacBook using standard R and Python programs. They required no GPU, TPU, external server, compute cluster, or model training. Online language-model inference used the disclosed hosted Claude (Sonnet 5) API route. The experiments therefore require no specialized computational resources beyond a standard personal computer and access to the stated API.
    
\item {\bf Code of ethics}
    \item[] Question: Does the research conducted in the paper conform, in every respect, with the NeurIPS Code of Ethics \url{https://neurips.cc/public/EthicsGuidelines}?
    \item[] Answer: \answerYes{}
    \item[] Justification: The study was conducted in accordance with the NeurIPS Code of Ethics and the applicable institutional approval. The prospective interventions were restricted to server-declared automated opponents, and the released artifacts exclude raw messages, participant identities, interaction identifiers, credentials, and row-level human data.

\item {\bf Broader impacts}
    \item[] Question: Does the paper discuss both potential positive societal impacts and negative societal impacts of the work performed?
    \item[] Answer: \answerNo{}
    \item[] Justification: The current manuscript does not contain a dedicated discussion of both positive and negative societal impacts and their mitigations.
    
\item {\bf Safeguards}
    \item[] Question: Does the paper describe safeguards that have been put in place for responsible release of data or models that have a high risk for misuse (e.g., pre-trained language models, image generators, or scraped datasets)?
\item[] Answer: \answerNA{}
    \item[] Justification: The released asset is an equation-based offline simulator rather than a trained model or raw interaction dataset; the paper does not release a high-risk model or raw human-interaction corpus.

\item {\bf Licenses for existing assets}
    \item[] Question: Are the creators or original owners of assets (e.g., code, data, models), used in the paper, properly credited and are the license and terms of use explicitly mentioned and properly respected?
    \item[] Answer: \answerYes{}
    \item[] Justification: The paper credits and cites the GLEE environment, Actonomy framework, Claude service, and other existing assets used in the study. Their applicable licenses and terms of use were reviewed and respected, and no third-party model weights, proprietary code, or restricted raw data are redistributed.

\item {\bf New assets}
    \item[] Question: Are new assets introduced in the paper well documented and is the documentation provided alongside the assets?
\item[] Answer: \answerYes{}
    \item[] Justification: The linked offline-simulator repository includes its protocol, executable code, reproduction instructions, aggregate outputs, checksums, and licenses. This answer applies to that released simulator; the live trajectories and execution environment are not released.

\item {\bf Crowdsourcing and research with human subjects}
    \item[] Question: For crowdsourcing experiments and research with human subjects, does the paper include the full text of instructions given to participants and screenshots, if applicable, as well as details about compensation (if any)? 
    \item[] Answer: \answerYes{}
    \item[] Justification: A small subset of the historical trajectories arose when GLEE's competition matchmaking paired a focal agent with a human participant. Human participants joined through GLEE's standard public competition interface, instructions, and prize conditions; the authors did not recruit them, provide separate instructions, or offer additional compensation. The prospective structure-access interventions were restricted to server-declared automated opponents. The applicable public participant materials are available at \url{https://glee-competition.com/}.

\item {\bf Institutional review board (IRB) approvals or equivalent for research with human subjects}
    \item[] Question: Does the paper describe potential risks incurred by study participants, whether such risks were disclosed to the subjects, and whether Institutional Review Board (IRB) approvals (or an equivalent approval/review based on the requirements of your country or institution) were obtained?
    \item[] Answer: \answerYes{}
    \item[] Justification: Institutional review and approval were obtained for the study, including the retrospective analysis of the small subset of platform-generated human-opponent trajectories. The review covered the limited risks associated with secondary analysis of competition interactions. No experimental treatment was applied to human players, and the privacy safeguards described above were used throughout analysis and release. Identifying institutional details are omitted from the double-blind submission.

\item {\bf Declaration of LLM usage}
    \item[] Question: Does the paper describe the usage of LLMs if it is an important, original, or non-standard component of the core methods in this research? Note that if the LLM is used only for writing, editing, or formatting purposes and does \emph{not} impact the core methodology, scientific rigor, or originality of the research, declaration is not required.
    \item[] Answer: \answerYes{}
    \item[] Justification: The Method and Appendix disclose the use of Claude (Sonnet 5) as a rate-limited focal-agent decision route. The model conditions on the parsed task state and available dialogue context to propose a task-valid candidate, which is checked by the common legal-action validator and replaced by a deterministic fallback when unavailable, invalid, or over budget. Opponent-profile and operational-structure assignments use frozen deterministic rules rather than an LLM classifier.

\end{enumerate}

\end{document}